\documentclass[usenatbib,usegraphicx]{mn2e}

\usepackage{graphicx}

\usepackage{times}
\usepackage{aas_macros}
\usepackage{pdfpages}
\usepackage{grffile}
\usepackage{amsmath}
\usepackage{amsfonts}
\usepackage{widetext}
\begin{document}

% write title (with email and institute)
\title[An improved method for polarimetric image restoration]{An improved method for polarimetric image restoration in interferometry}
\author[L. Pratley and M. Johnston-Hollitt]
{Luke Pratley$^1$\thanks{E-mail: Luke.Pratley@gmail.com, Melanie.Johnston-Hollitt@gmail.com}, Melanie Johnston-Hollitt$^{1}$\\
$^1$ School of Chemical \& Physical Sciences, Victoria University of Wellington, PO Box 600, Wellington 6140, New Zealand}
\maketitle

\begin{abstract}
Interferometric radio astronomy data require the effects of limited coverage in the Fourier plane to be accounted for via a deconvolution process. For the last 40 years this process, known as `cleaning', has been performed almost exclusively on all Stokes parameters individually as if they were independent scalar images. However, here we demonstrate for the case of the linear polarisation $\mathcal{P}$, this approach fails to properly account for the complex vector nature resulting in a process which is dependant on the axis under which the deconvolution is performed. We present here an improved method, `Generalised Complex CLEAN', which properly accounts for the complex vector nature of polarised emission and is invariant under rotations of the deconvolution axis.  We use two Australia Telescope Compact Array datasets to test standard and complex CLEAN versions of the H\"{o}gbom and SDI CLEAN algorithms. We show that in general the Complex CLEAN version of each algorithm produces more accurate clean components with fewer spurious detections and lower computation cost due to reduced iterations than the current methods. In particular we find that the Complex SDI CLEAN produces the best results for diffuse polarised sources as compared with standard CLEAN algorithms and other Complex CLEAN algorithms. Given the move to widefield, high resolution polarimetric imaging with future telescopes such as the Square Kilometre Array, we suggest that Generalised Complex CLEAN should be adopted as the deconvolution method for all future polarimetric surveys and in particular that the complex version of a SDI CLEAN should be used. 
\end{abstract}

% look up list of allowable keywords for this section

\section{INTRODUCTION}
\label{sec:intro}
The challenge of deconvolution is wide-spread in radio astronomy, ranging from aperture synthesis \citep{R&H1960} to rotation measure synthesis \citep{bre05}.  In particular, the technique of aperture synthesis allows observers to use radio interferometers to observe the sky at higher resolution and sensitivity than possible with the largest individual radio telescopes. Such advantages provide the motivation for the next-generation interferometric radio telescopes, such as the LOw Frequency ARray \citep[LOFAR;][]{vH13}, the Murchison Widefield Array \citep[MWA;][]{tingay13}, the Australian Square Kilometre Array Pathfinder \citep[ASKAP;][]{hotan14} and the Square Kilometre Array (SKA). However, to accurately interpret the observations of a radio interferometer, a method of image reconstruction needs to be applied. Image reconstruction in radio interferometry is often referred to as deconvolution, because it is used to deconvolve the effects of limited uv-coverage on the observation.

A number of possible methods can be used to solve the deconvolution problem including CLEAN, maximum entropy and compressed sensing. Maximum entropy methods are in particularly useful for deconvolution of mosaic images \citep{cor85,cor88}. Additionally modern methods of deconvolution, such as compressed sensing, have begun to be investigated for next-generation radio interferometers \citep{li11a,car14,gar15}, however they are still in their infancy and currently limited to very narrow observational bandwidths. 

Historically, the CLEAN algorithm and its variations have been successfully applied as a standard method of deconvolution in radio interferometry. The CLEAN algorithm was first introduced by \cite{hog74}, and is commonly referred to as H\"{o}gbom CLEAN, and presents a statistically correct least-squares fit of the measured $uv$-data \citep{sch78}. Since then, variations of the CLEAN method have been developed to improve the speed and quality of the algorithm \citep{cla80,sch84}. Furthermore, variations have been developed to improve deconvolution of resolved and extended structures, such as SDI (Steer-Dwedney-Ito) CLEAN \citep{ste84} and Multi-Scale CLEAN \citep{cor08}. With the recent introduction of next-generation radio telescopes with non-coplanar baselines (LOFAR, MWA), wide-field imaging variations of CLEAN have also been developed \citep{off14}.

One of the strengths of radio astronomy is that polarisation can be measured accurately at radio wavelengths. This enables science that is not accessible at other wavelengths, such as the study of cosmic magnetism which is a key driver of next-generation telescopes like the SKA \citep{mjh15}. Additionally, the observed polarisation is vital to calibration, which can effect image quality in total intensity images.

To date, the main motivation for deconvolution and for the CLEAN algorithm has been the deconvolution of total intensity (Stokes $\mathcal{I}$) images. In practice, the same CLEAN algorithms are applied to both the intensity and the polarimetric image, Stokes $\mathcal{Q}$, $\mathcal{U}$, and $\mathcal{V}$ which respectively represent the two linear and one circular polarisations. The polarised Stokes images themselves are intermediate products produced to extract either the total linear polarisation or the total polarisation which are complex valued vector quantities derived from the component Stokes parameters. In the majority of cases, celestial sources have no detectable circular polarisation, while detectable linear polarisation is common. Thus, of particular interest is the total linear polarisation, $\mathcal{P}$, which is given by:
\begin{equation}
\label{eq:linearpol}
\mathcal{P}=\mathcal{Q}+i\mathcal{U}\, .\\
\end{equation}
 
Because deconvolution has been motivated for Stokes $\mathcal{I}$, the Stokes $\mathcal{Q}$ and $\mathcal{U}$ images are typically deconvolved individually, rather than together as a complex linear polarisation image as suggested by Equation \ref{eq:linearpol}. As a result the CLEAN methods used for the past 40 years do not correctly account for the complex vector nature of $\mathcal{P}$ \footnote{Several older papers within the VLBI literature mention the task CXCLN \citep{cot84,cot92}, in classic AIPS, which uses a H\"{o}gbom algorithm described as a `complex CLEAN'. This task is functionally different, since it was developed to account for asymmetric uv-coverage of Stokes $\mathcal{Q}$ and $\mathcal{U}$ in VLBI observations when interferometers using circular feeds have only measured one of the circular correlations LR or RL \citep{cot92}. The understanding of the VLBI community is that this approach is only needed when only half the correlations are measured, and that it is identical to the traditional approach in cases where all 4 correlations (RR, RL, LR and LL) are measured \citep{aar97}, but as we will show here, this is incorrect. Note we have adopted the name `Generalised Complex CLEAN' for our method to avoid confusion with CXCLN.}.

All observations with interferometers must be referenced to a standard coordinate system, which can be either fixed with respect to some external coordinate system (basis) or derived from the physical orientation of the crossed dipoles or circular feed horns as they are either fixed on the Earth (e.g. for the MWA), or in the case of a dish array, when the instrument is stowed (e.g. for the Australia Telescope Compact Array and Jansky Very Large Array for linear or circular feeds, respectively). The position angle of the polarisation, $\chi$, of an electric field is measure by an interferometer as components of the two linear polarisations Stokes $\mathcal{Q}$ and $\mathcal{U}$ made with respect to the coordinate axis such that:
\begin{equation}
\mathcal{\chi}=\frac{1}{2}\rm{atan}\frac{\mathcal{U}}{\mathcal{Q}}\, .\\
\end{equation}

For a given polarisation position angle, the magnitude of the measured Stokes $\mathcal{Q}$ and $\mathcal{U}$ components will change when measured in a rotated coordinate system. However, the final complex polarisation vector reconstructed from the measured linear polarisations is rotationally invariant. The physical linear polarisation vector as a geometric object does not depend on the coordinate system used to measure it. For this reason, rotation of the dipoles or circular feeds with respect to the source during Earth rotation synthesis must be corrected and transformed back to the original coordinate system during calibration via introduction of a phase rotation term \footnote{This will be the case for all antennas unless the instrument has triaxial feeds such as ASKAP \citep{hotan14} which keep the dipoles at a constant position relative to the source.}. Having taken care to correctly account for the rotation of the dipole or circular feed as sources are tracked across the sky, it follows that one should apply a deconvolution process that produces results which are rotationally invariant. The observation of linear polarisation vectors restored from the deconvolution process should also be geometric objects that are independent of coordinate system. Otherwise, the linear polarization images may depend on the choice of coordinate systems.  The standard CLEAN methods are not invariant under rotations of $\mathcal{P}$ when $\mathcal{Q}$ and $\mathcal{U}$ are CLEANed separately, and the final sky model constructed by standard CLEAN algorithms depends on the chosen orientation of the $\mathcal{Q}$ and $\mathcal{U}$ axes. In this work, we demonstrate that the CLEANed model will depend on the orientation chosen to perform the deconvolution, then provide an alternative CLEAN deconvolution method that constructs a model independent of the orientation of the $\mathcal{Q}$ and $\mathcal{U}$ axes. 

The alternative method is called `Generalised Complex CLEAN', and is a modification of the CLEAN algorithm. Generalised Complex CLEAN works by locating residual peaks in linear polarisation, rather than locating the peaks independently in Stokes $\mathcal{Q}$ and $\mathcal{U}$. Once the peaks are found, an appropriate model is subtracted according to the CLEAN algorithm variant. Here we demonstrate that the deconvolution process is invariant under rotations of $\mathcal{P}$, meaning that the residuals do not depend on the chosen axis for $\mathcal{Q}$ and $\mathcal{U}$. We then show that complex CLEANing $\mathcal{P}$ will provide an advantage over traditional CLEANing by detecting more CLEAN components in sources near the noise level, whilst simultaneously detecting less spurious components scattered across the image.

The paper is arranged as follows: first we show using high quality Australia Telescope Compact Array (ATCA) observations that the model constructed by CLEANing $\mathcal{Q}$ and $\mathcal{U}$ independently is biased to the chosen axes for deconvolution. Then we introduce the Complex H\"{o}gbom and SDI CLEAN methods, and describe how they can be applied to deconvolve images of complex linear polarisation $\mathcal{P}$. Finally, we then demonstrate that complex CLEANing produces a CLEAN model that does not depend on the choice of axes used for deconvolution and show that it produces superior results to current methods.

By showing that the H\"{o}gbom and SDI CLEAN methods can be extended to not depend on the choice of axes used for deconvolution, we set the stage for extending other methods of deconvolution from real valued images to complex valued images.

\section{Observational Data}
In this work, we use two ATCA observations to compare the effects of using a standard and Generalised Complex CLEAN on synthesized linear polarisation images. In this section, we present the details of the two observations.

\begin{figure}
\centering
        \includegraphics[width=9cm, trim=0 0 0 1.3cm]{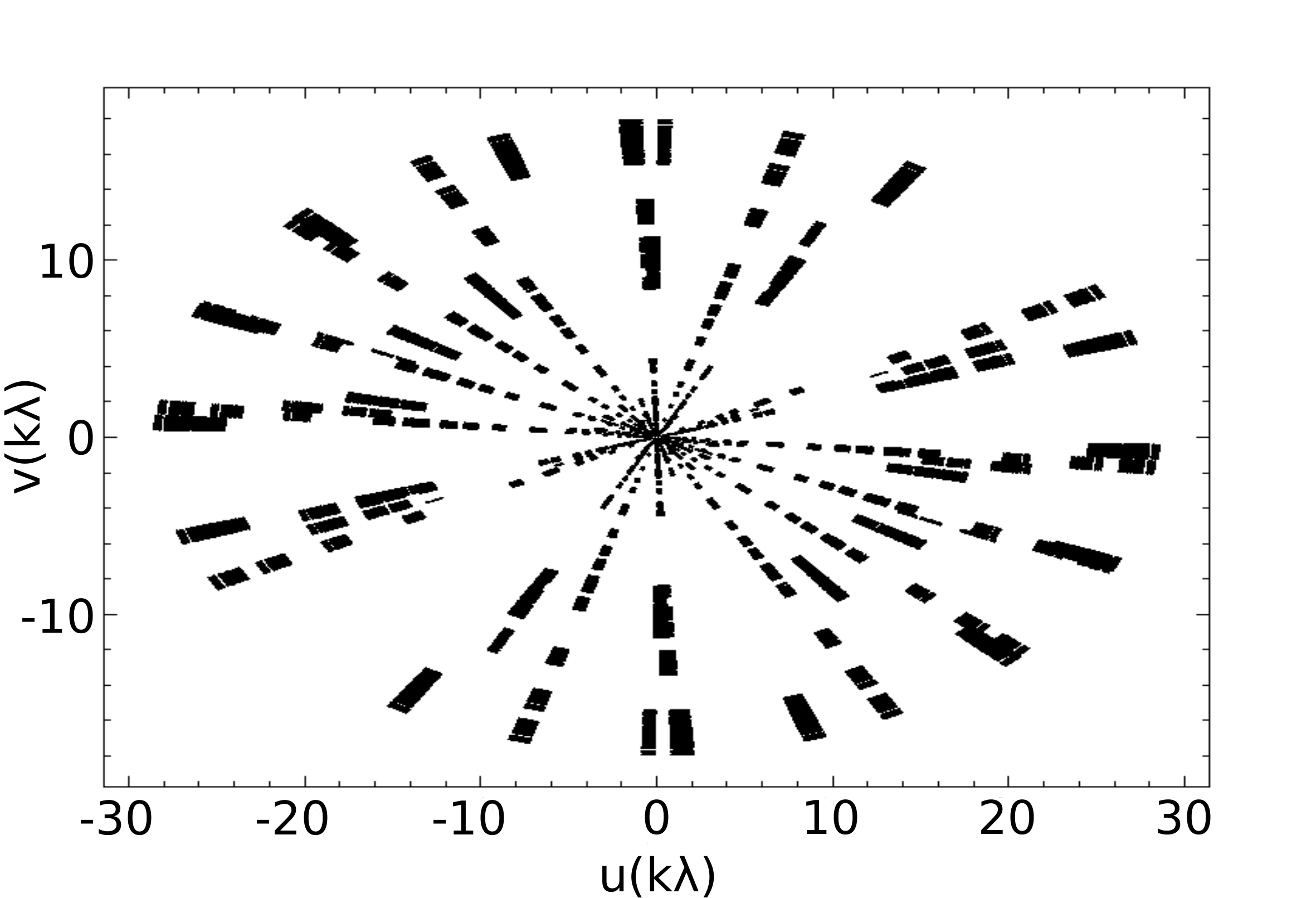} 
    \caption{UV coverage of PKS J0334-3900 in units of kilo-$\lambda$. Note that this is the uv-coverage based on the full bandwidth.}
    \label{fig:0334uv}
\end{figure}

\begin{figure}
\centering
    \includegraphics[width=0.5\textwidth]{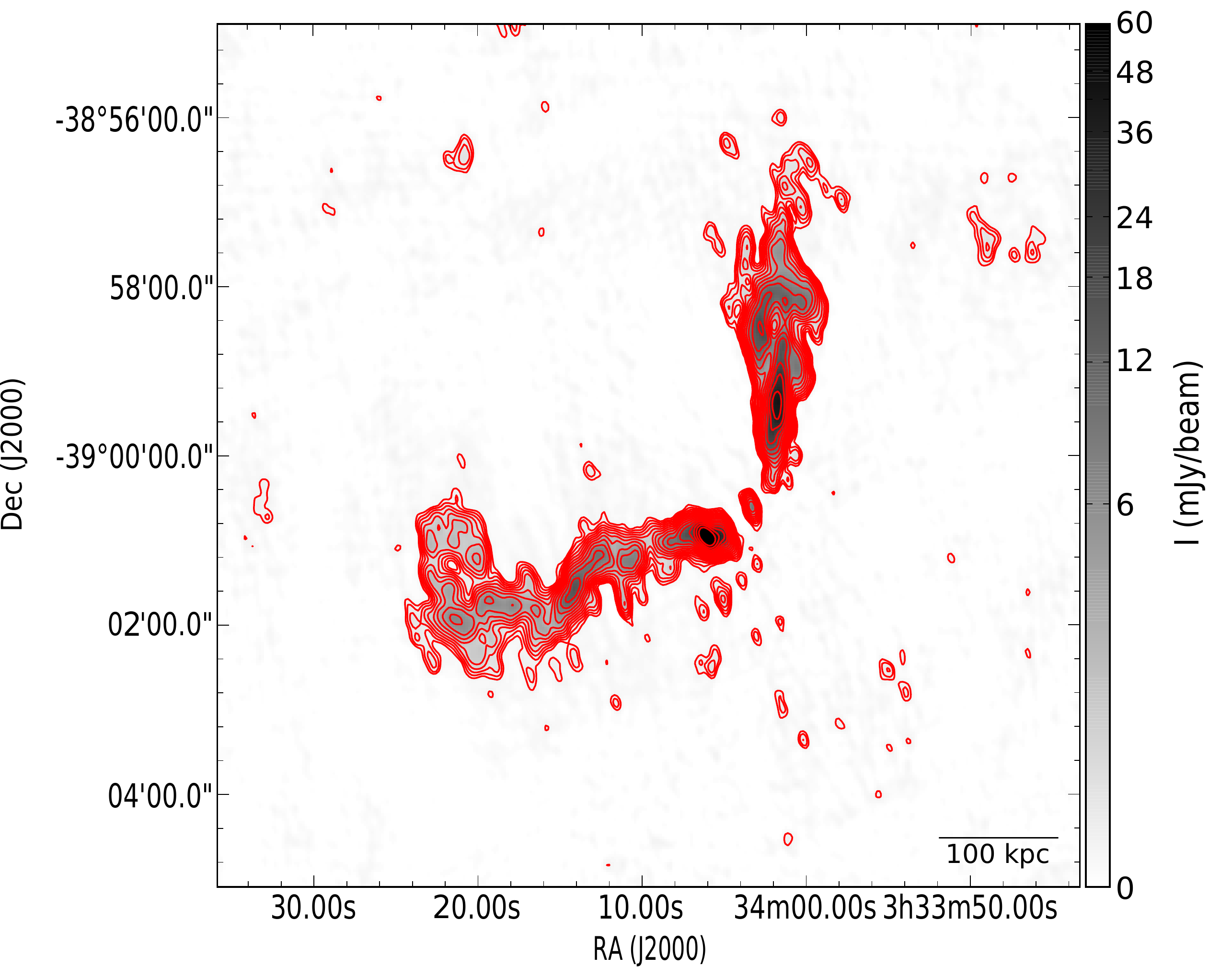}
    \caption{1.4 GHz image of PKS J0334-3900 used in this work. The RMS noise of the Stokes $\mathcal{I}$ image is 140 $\mu$Jy b$^{-1}$ and the beam is 1.2''$\times$5.8'', pa = 11.6$^\circ$.}
    \label{fig:0334StokesI}
\end{figure}

The first data set we use is a 1.4 GHz observation of the radio galaxy PKS J0334-3900 located in the galaxy cluster Abell 3135. PKS J0334-3900 is an intermediate FR I/II radio galaxy with diffuse linearly polarised jets classified as a bent tailed source. The full details of the observations and polarisation analysis can be found in our previously published work \citep{pra13} and the observations used here are given in Table \ref{tab:obsinfo}. However, since that work, the data has been reprocessed using a standard self-calibration routine, improving the sensitivity of the 1.4 GHz images, and it is these reprocessed data that are used here. As before the synthesized images were made using uniform weighting which results in a synthesized beam of 1.2 x 5.8 arcsec. A plot of the $uv$ coverage can be found in Figure \ref{fig:0334uv} and Figure \ref{fig:0334StokesI} shows the 1.4 GHz image. 

\begin{table}
\center
\caption{Observation time per baseline configuration, and RMS noise in the final Stokes $\mathcal{Q}$ and $\mathcal{U}$ images generated from co-added uv-data from all available configurations.}
\label{tab:obsinfo}
\begin{tabular}{l c c c c}
\hline
Source & Config. & Time & RMS noise & Beam\\
 & & (min) & ($\mu$Jy/beam)& (arcsec, deg.)\\
\hline
PKS J0334-3900 & 6A& 59& 160 & 1.15$\times$5.82, 11.62$^\circ$ \\
			& 1.5A& 76&\\
			& 750A& 79.7&\\
			& 375&  75.4&\\
PKS B1637-771 & 6A& 589.8& 35 &  5.47$\times$5.05, -36.73$^\circ$\\
		     & 6B& 571.2&\\
		     & 6C& 637.8&\\
		     & 6E& 602.4&\\
		    & 750D& 723.6&\\
\hline
\end{tabular}
\end{table}

\begin{figure}
\centering
    \includegraphics[width=9cm, trim= 0 0 0 1.3cm]{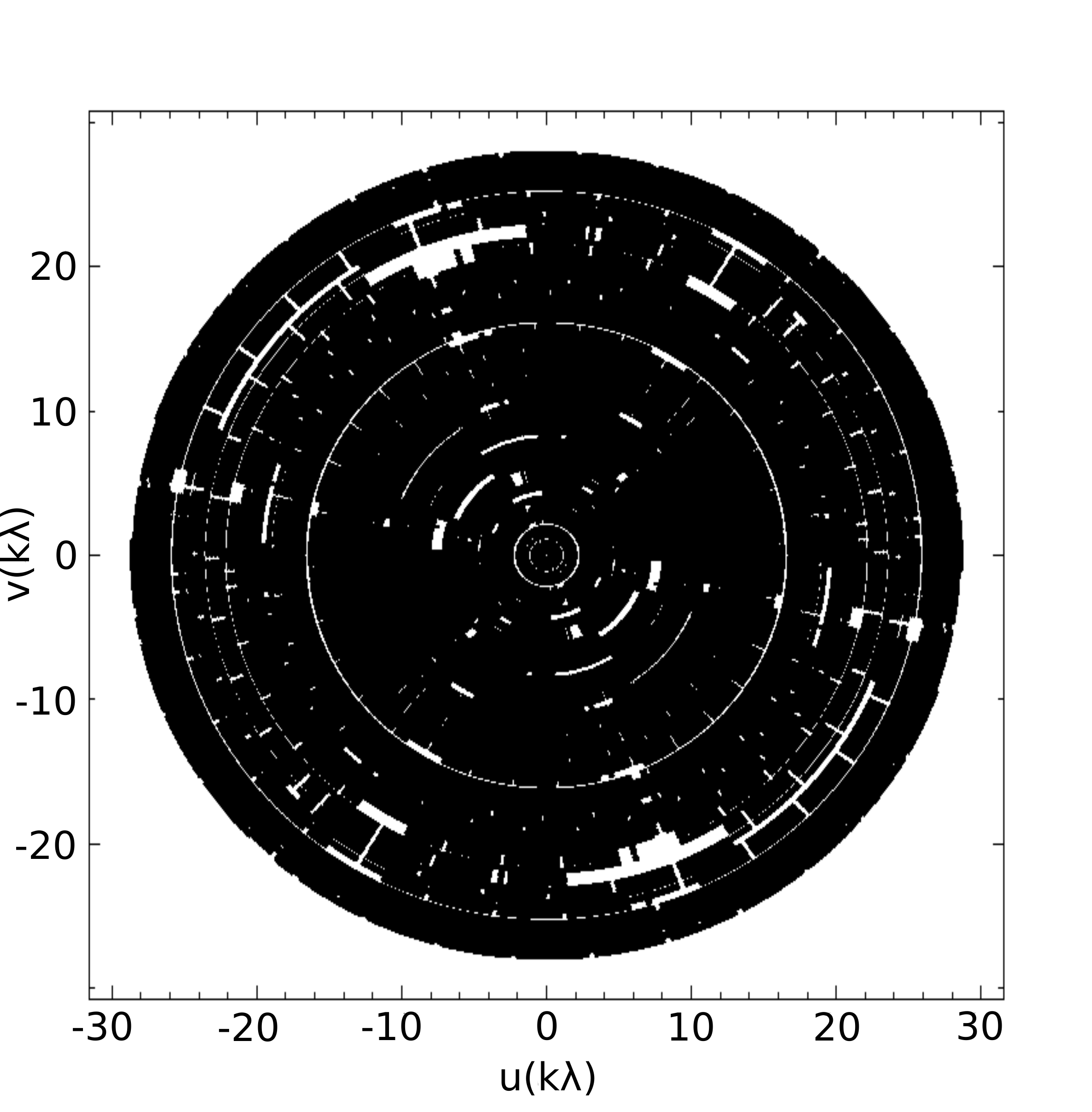} 
    \caption{UV coverage of PKS B1637-771 in units of kilo-$\lambda$. Note that this is the uv-coverage based on the full bandwidth.}
    \label{fig:1637-77uv}
\end{figure}

\begin{figure}
\vspace{-0.08cm}
\centering
    \includegraphics[width=0.46\textwidth, trim=1.0cm 0 1.0cm 0cm]{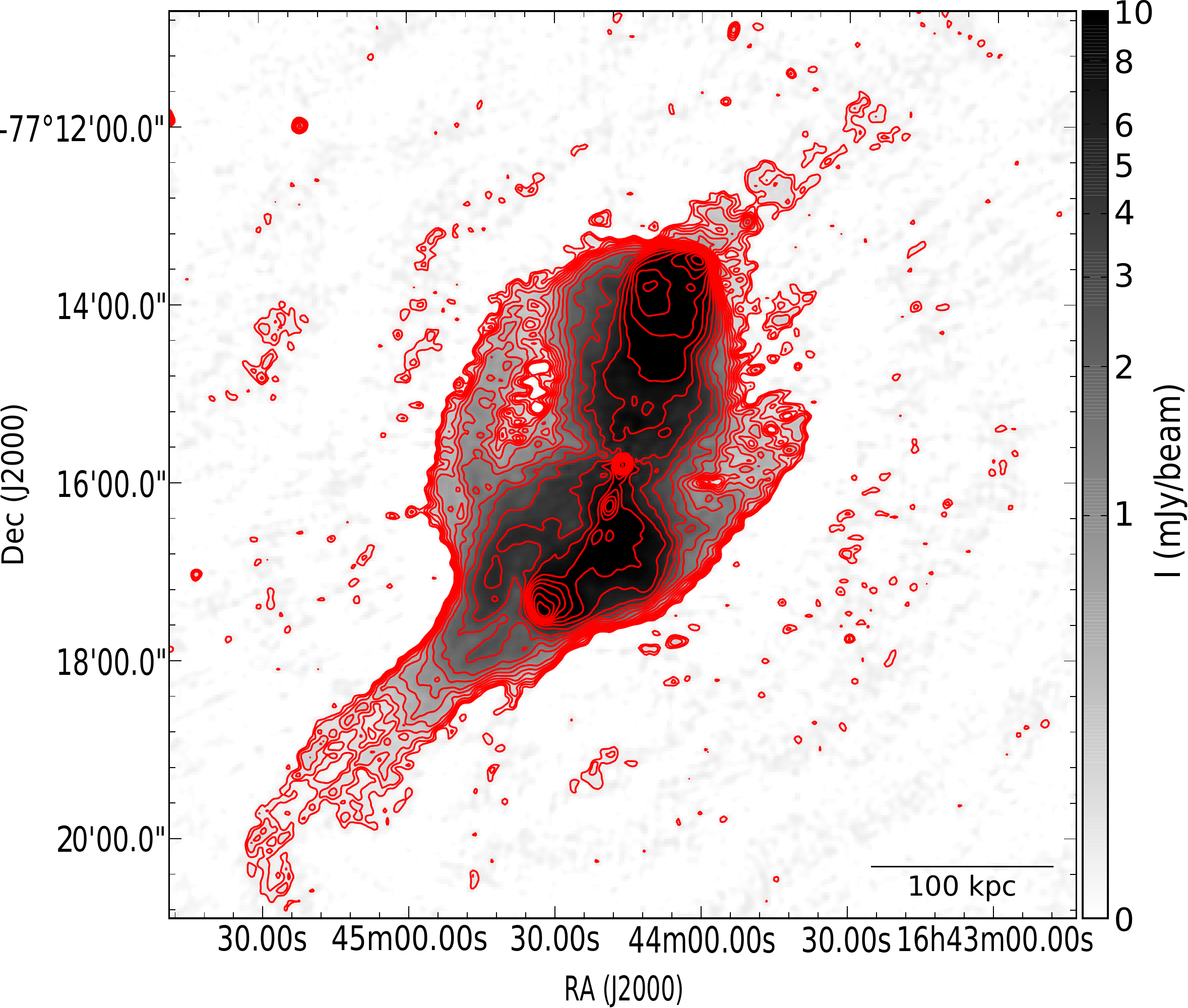}
    \caption{1.4 GHz image of PKS B1637-771 used in this work. The RMS noise of the Stokes $\mathcal{I}$ image is 35 $\mu$Jy b$^{-1}$ and the beam is approximately circular and 5'' in diameter.}
    \label{fig:1637StokesI}
\end{figure}

The second dataset is from an observation of the radio galaxy PKS B1637-771 at 1.4 GHz. PKS B1637-771 is a diffuse double lobed radio galaxy, previously stated to be an FRII galaxy \citep{Sadler14}. A low-surface brightness cocoon of emission surrounds the jets of the radio galaxy, and the complex wispy structure of the cocoon suggests some disruption in the environment or jets. As the jets and cocoon are linearly polarised, this is an excellent example of a highly complex source with diffuse polarised emission and is thus a good case on which to test polarisation cleaning methods. 

All ATCA observations of PKS B1637-771 were extracted from the archive (PI: Leahy, project code C888). They were performed between 2000 and 2001 using a phase centre of RA$_{\rm J2000}$ = 16h 44m 20.00s, Dec$_{\rm J2000}$ = -77$^\circ$ 15' 30.00''. The observations were centred at at frequency of 1384 MHz with a bandwidth of 128 MHz over 32 channels. The channels on the edge of the band were removed, leaving an effective bandwidth of 96 MHz. Due to self-generated interference in the 1384 MHz observation, the 8 MHz interval centred at 1408 MHz was also removed.

The observation used five baseline configurations (see Table \ref{tab:obsinfo}), providing excellent UV-coverage. Taking all configurations into account, the total integration time was 50.78 hours (3046.8 minutes). The flux density scale was set relative to the unresolved calibrator PKS B1934-638, using an assumed total flux density at 1384 MHz of 14.94 $\pm$ 0.01 \citep{reynolds94}. The time variations in complex antenna gains and bandpass were calibrated using observations of the unresolved source PKS B1718-649. The observations were calibrated using MIRIAD, while following the standard ATCA pre-CABB reduction method \citep{sault95}. Synthesized images were made using uniform weighting of the calibrated data, resulting in a synthesized beam with full width half maximum of 5.5 x 5.0 arcsec. A plot of the $uv$ coverage can be found in Figure \ref{fig:1637-77uv} and Figure \ref{fig:1637StokesI} shows the final Stokes $\mathcal{I}$ image.

The $\mathcal{Q}$ and $\mathcal{U}$ root-mean-squared (RMS) noise for each observation can be found in Table \ref{tab:obsinfo}.

The data from PKS J0334-3900 were used to compare the standard and complex H\"{o}gbom CLEANs, whilst the observation of PKS B1637-771 was used to compare both the standard and complex SDI CLEANs. In the case of PKS J0334-3900 the standard H\"{o}gbom CLEAN on the polarised data was found to produce publishable images \citep[e.g.][]{pra13}, however for PKS B1637-771 H\"{o}gbom CLEAN produces a low quality deconvolution of the observation for this diffuse source, and is plagued by artefacts illustrating the well known poor performance of H\"{o}gbom CLEAN on low surface brightness, complex emission. For such sources an SDI CLEAN performs better when inspecting the residuals, and has no visible artefacts. The need to accurately CLEAN diffuse sources partly motivates the need for developing the complex SDI CLEAN method.

\begin{table}
\caption{List of symbols.}
\begin{tabular}{l l}
\hline
Symbol & Description \\
\hline
$\mathcal{I}$, $\mathcal{Q}$, $\mathcal{U}$, $\mathcal{V}$ & Stokes parameters \\
$\mathcal{P}$ & Linear polarisation, $\mathcal{P}=\mathcal{Q}+i\mathcal{U}$ \\
$\mathcal{B}_D$ & Synthesized beam, point spread function \\
$\mathcal{I}_D,\mathcal{Q}_D,\mathcal{U}_D$ & Synthesized image of respective intensity \\
$\mathcal{P}_D$ & Linear polarisation synthesized image, $\mathcal{P}_D=\mathcal{Q}_D+i\mathcal{U}_D$ \\
$\mathcal{N}_\mathcal{Q},\mathcal{N}_\mathcal{U}$ & Stokes $\mathcal{Q}$ and $\mathcal{U}$ Gaussian noise\\
$\mathcal{C}_\mathcal{Q},\mathcal{C}_\mathcal{U}$ & Stokes $\mathcal{Q}$ and $\mathcal{U}$ CLEAN component images, model images\\
$\mathcal{R}_\mathcal{Q},\mathcal{R}_\mathcal{U}$ & Stokes $\mathcal{Q}$ and $\mathcal{U}$ residual images\\
$\mathcal{R}_{\mathcal{Q}_F},\mathcal{R}_{\mathcal{U}_F}$ & Stokes $\mathcal{Q}$ and $\mathcal{U}$ final residual images\\
$\mathcal{R}_{\mathcal{P}_F}$ & Linear polarisation final residual image\\
$\mathcal{C}_\mathcal{P}$ & Linear polarisation CLEAN component image, model image\\
$\mathcal{M}$ & Model generated from SDI CLEAN clip.\\
$\mathcal{M}_D$ & Synthesized model $\mathcal{M}_D=\mathcal{M}\star\mathcal{B}_D$.\\
$\gamma$ & Gain used in CLEAN \\
$\alpha$ & Clip value used in SDI CLEAN\\
$\eta$ & Damping factor used in SDI CLEAN\\
$\sigma$ & Standard deviation of Gaussian noise in Stokes $\mathcal{Q}$ and $\mathcal{U}$\\
$\mu_\mathcal{P}$ & Image of average complex linear polarisation $\mathcal{P}$\\
$\xi$ & Constraint used as a cutoff for CLEAN, i.e. $\xi = 3\sigma$\\
$(\mathcal{Q},\mathcal{U})$ & Chosen coordinate axes/frame for Stokes $\mathcal{Q}$  and $\mathcal{U}$ \\
$(\mathcal{Q}^\prime$, $\mathcal{U}^\prime)$ & Linear Stokes parameters represented in a rotated frame.\\
\hline
\label{tab:symbols}
\end{tabular}
\end{table}

\section{Standard CLEAN method applied to linear polarisation data}
\label{sec:standard}
The standard clean procedure picks the highest peaks in the $\mathcal{Q}$ and $\mathcal{U}$ synthesized images $\mathcal{Q}_D$ and $\mathcal{U}_D$, respectively, and then iteratively removes the synthesized beam associated with those peaks via deconvolution. This is the deconvolution used in all major radio astronomy imaging software including AIPS, Newstar \citep{ike01}, MIRIAD \citep{sault95}, CASA \citep{mcmullin07}, WSClean \citep{off14} and the pipelines based on these packages for instruments such as LOFAR, MWA and ASKAP. Even in more recent packages such as CASA and WSClean where there is the ability to deconvolve in different ways, the standard method of searching for peaks in $\mathcal{Q}$ and $\mathcal{U}$ independently in $\mathcal{Q}_D$ and $\mathcal{U}_D$, respectively, is most often used. \footnote{CASA allows three options beyond the standard cleaning approach when using a Clark CLEAN only: i) searching for peaks in $\mathcal{I}$, ii) searching for peaks in $\mathcal{I}$ and $\mathcal{V}$ simultaneously, or iii) $\mathcal{I}$ and total polarisation $\sqrt(\mathcal{Q}^2 + \mathcal{U}^2 + \mathcal{V}^2)$ simultaneously. In all cases it then produces individual clean component images for each Stokes parameter. This latter approach is of most relevance here but this is designed to constrain peaks so as to select the most highly polarised components associated with a continuum (Stokes $\mathcal{I}$) source first. Constraining components to have both a strong peak in $\mathcal{I}$ and total polarisation is in fact worse for some science applications than just searching for peaks in $\mathcal{P}$  e.g. when looking at diffuse polarisation in the Galactic Plane. In such cases the spatial scales of emission in Stokes $\mathcal{I}$ and polarisation is different and the interferometer can resolve out the Stokes $\mathcal{I}$ whilst still detecting polarised structures resulting in strong polarised signals with no matching continuum emission (see \citealp{gae01, deB05} for examples). For this reason LOFAR which runs AWImager only cleans in the Stokes parameters independently. The most flexible current CLEANing package is WSClean \citep{off14} which will allow searches for peaks in the sum of squares of any sensible combination of polarisations, without requiring a Stokes $\mathcal{I}$ counterpart. While WSClean can provide this functionality through its ``jointpolarizations" option, there are some issues when also implementing multi-scale clean and cleaning jointly in $\mathcal{Q}$ and $\mathcal{U}$  is not usually how the software is run.}

However, $\mathcal{Q}$ and $\mathcal{U}$ are treated as individual scalar images, which doesn't account for the vector nature of linear polarisation, meaning that the arbitrarily chosen linear polarisation axes directions will be preferentially CLEANed. Here we will lay out the mathematical foundations for the CLEANing process and demonstrate its rotational dependance on the two datasets mentioned above by rotating the coordinate system before CLEANing.

First we define a standard set of symbols in Table \ref{tab:symbols}. In this work, the hypothesis is that  deconvolution process, represented by the function $\mathbb{D}$, is not invariant under rotations $R(\theta_0)$ of linear polarisation,
\begin{equation}
R(\theta_0)=\begin{bmatrix}\cos(\theta_0) & -\sin(\theta_0) \\ \sin(\theta_0) & \cos(\theta_0) \end{bmatrix}\, .
\end{equation}
 Saying that $\mathbb{D}$ is rotationally invariant, is the same as the mathematical statement
\begin{equation}
\mathbb{D}\left [\mathcal{P}_D\right] \equiv R(-\theta_0)\mathbb{D}\left [R(\theta_0)\mathcal{P}_D\right]\, ,
\end{equation}
being true for all angles $\theta_0$.

If we define linear polarisation in a rotated frame as
\begin{equation}
\mathcal{P}^\prime (x,y,\theta_0) = R(\theta_0)\begin{bmatrix}\mathcal{Q}(x, y) \\ \mathcal{U}(x, y) \end{bmatrix}\, ,
\end{equation}
then a rotationally invariant CLEAN method would provide CLEAN components that can be related to the non-rotated frame by a simple rotation
\begin{equation}
C_{\mathcal{P}}=R(-\theta_0)C_{\mathcal{P}^\prime}\,.
\end{equation}
However, if CLEAN is not rotationally invariant, then 
\begin{equation}
C_{\mathcal{P}}\neq R(-\theta_0)C_{\mathcal{P}^\prime}\,.
\end{equation}
Rotational invariance of CLEAN is critical for physical results. For instance, the orientation of the dipole could be used as the reference frame of $\mathcal{Q}$ and $\mathcal{U}$, if CLEAN is not rotationally invariant then rotating the dipole of the antenna could produce different scientific results. 

\begin{figure*}
\centering
\begin{minipage}[c]{\textwidth}
\centering
    \includegraphics[width=0.5\textwidth]{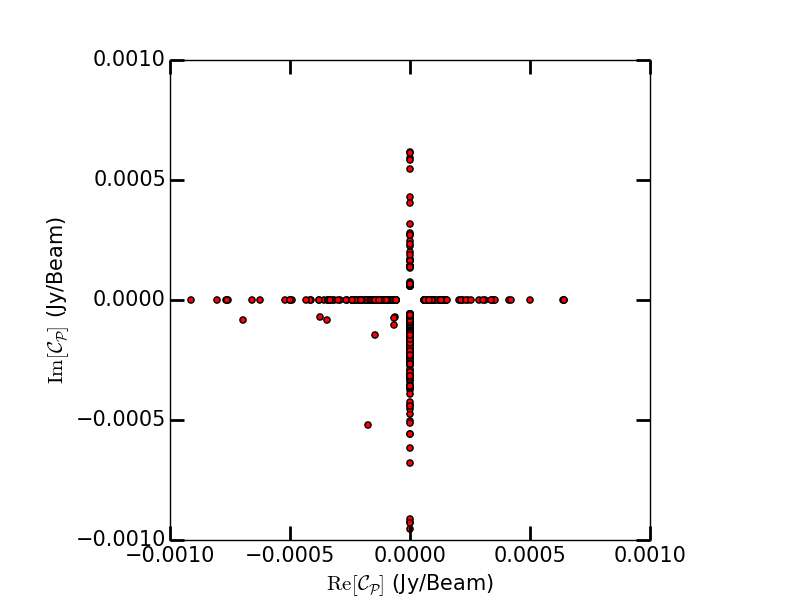}\includegraphics[width=0.5\textwidth]{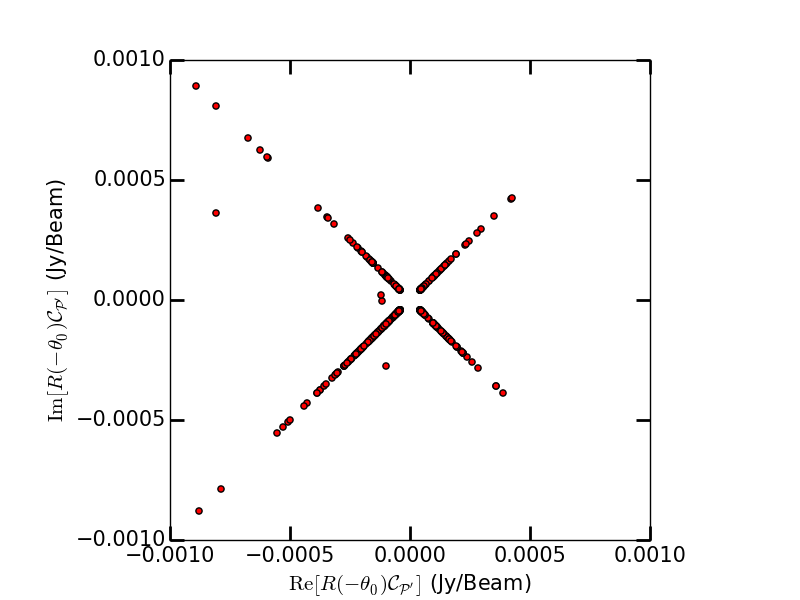}
    \caption{Plots of the linear polarisation states of the CLEAN components for PKS J0334-3900, generated from H\"{o}gbom CLEANing Stokes $\mathcal{Q}$ and $\mathcal{U}$ separately. Left: Polarisation states when the H\"{o}gbom CLEAN components are generated in the $\mathcal{Q}$ and $\mathcal{U}$ frame. Right: Polarisation states when the H\"{o}gbom CLEAN components are generated in the $\mathcal{Q}^\prime$ and $\mathcal{U}^\prime$ frame, where $\mathcal{Q}^\prime$ is the frame rotated by $\theta_0=45^\circ$ from the $\mathcal{Q}$ axis and then rotated back to the original coordinate frame. In each plot, the polarisation distribution of the model is biased to lie along the chosen $\mathcal{Q}$ and $\mathcal{U}$ axes used for deconvolution.}
    \label{fig:hogbomplot}
\end{minipage}
\end{figure*}

\begin{figure*}
\centering
\begin{minipage}[c]{\textwidth}
\centering
    \includegraphics[width=0.5\textwidth]{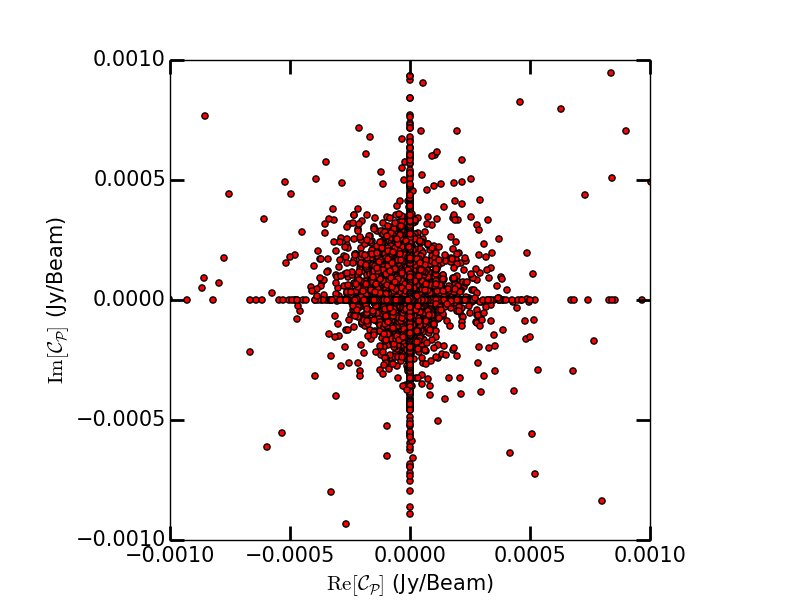}\includegraphics[width=0.5\textwidth]{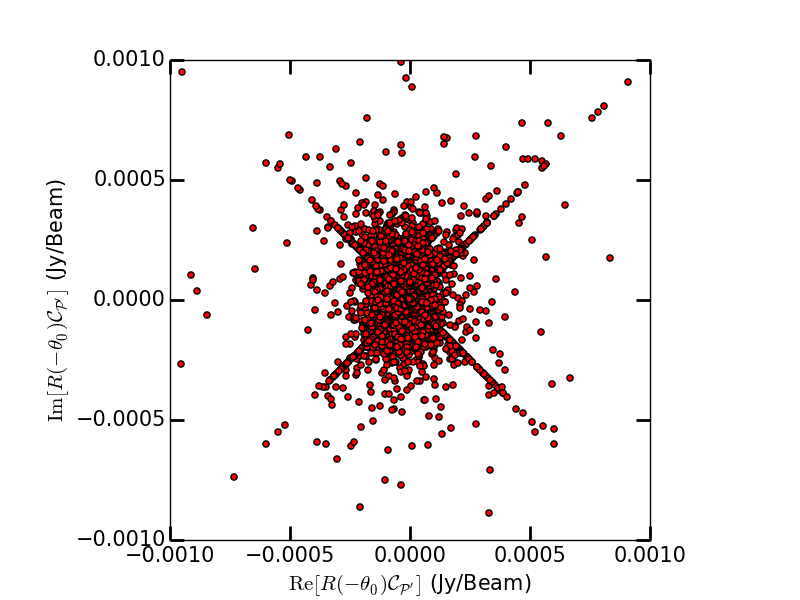}
    \caption{Plots of the linear polarisation states of the CLEAN components for PKS B1637-771, generated from SDI CLEANing Stokes $\mathcal{Q}$ and $\mathcal{U}$ separately. Left: Polarisation states when the SDI CLEAN components are generated in the $\mathcal{Q}$ and $\mathcal{U}$ frame. Right: Polarisation states when the SDI CLEAN components are generated in the $\mathcal{Q}^\prime$ and $\mathcal{U}^\prime$ frame, where $\mathcal{Q}^\prime$ is rotated by $\theta_0=45^\circ$ from the $\mathcal{Q}$ axis and then rotated back to the original coordinate frame. In each plot, the polarisation distribution of the model is biased by the chosen $\mathcal{Q}$ and $\mathcal{U}$ used for deconvolution.}
    \label{fig:steerplot}
\end{minipage}
\end{figure*}

Next, we present the effects of deconvolving synthesized Stokes $\mathcal{Q}_D$ and $\mathcal{U}_D$ images individually using standard H\"{o}gbom and SDI CLEANs. The images were CLEANed until the absolute maximum residual in $\mathcal{Q}$ and $\mathcal{U}$ reached the cutoff of 4 times the root-mean-squared (RMS) noise level. %We show that CLEANing dirty Stokes $\mathcal{Q}_D$ and $\mathcal{U}_D$ maps individually will bias the CLEAN components to lie along the the chosen $\mathcal{Q}$ and $\mathcal{U}$ axes used for deconvolution. 
To analyse the CLEAN component images for Stokes $\mathcal{Q}$ and $\mathcal{U}$, the polarisation vectors of each component in the model are plotted in a $\mathcal{Q}$$\mathcal{U}$ plane.  Plots are shown in Figures \ref{fig:hogbomplot} \& \ref{fig:steerplot}, showing the linear polarisation distribution of the CLEAN components for the original and rotated frames for a H\"{o}gbom CLEAN of the PKS J0334-3900 dataset and an SDI CLEAN of the PKS B1637-771 dataset, respectively. For the rotated frames (right panels of Figures \ref{fig:hogbomplot} \& \ref{fig:steerplot}) the deconvolution process is performed in the $\mathcal{Q}^\prime$ and $\mathcal{U}^\prime$ frame, where $\theta_0 = 45^\circ$ and then after deconvolution, the CLEAN components in the $\mathcal{Q}^\prime$ and $\mathcal{U}^\prime$ frame are rotated back to the original frame. If the CLEAN processes tested here were rotationally invariant both the original (left) and rotated (right) panels of Figures \ref{fig:hogbomplot} and \ref{fig:steerplot} would be identical. However, we see from these figures that polarisation distribution of the model is biased to lie along the chosen $\mathcal{Q}$ or $\mathcal{U}$ axis used for deconvolution and that $\mathcal{C}_\mathcal{P}\neq R(-\theta_0)\mathcal{C}_{\mathcal{P}^\prime}$. Thus, the standard CLEAN methods are not invariant under rotations of linear polarisation and as a result CLEANing using the present methods is not optimal for selecting the most physically important structures.

\begin{figure*}
\centering
\begin{minipage}[c]{\textwidth}
\centering
    \includegraphics[width=0.5\textwidth]{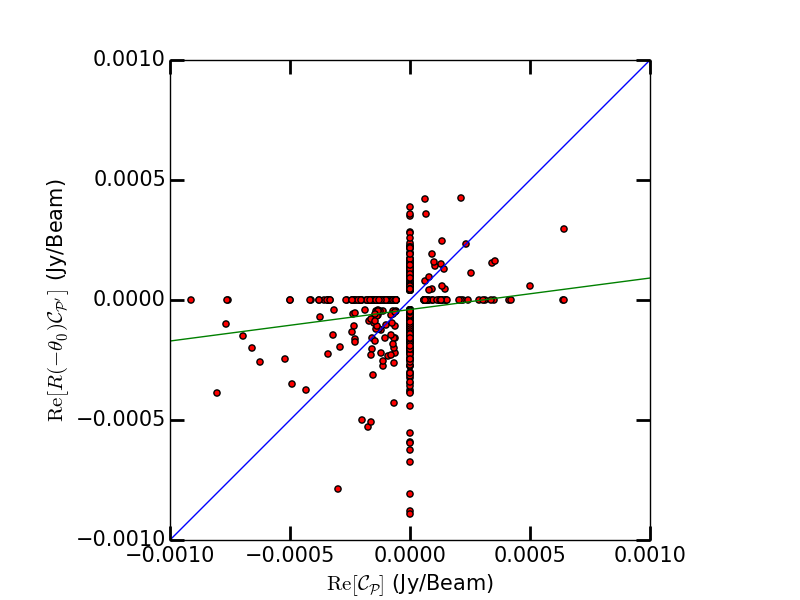}\includegraphics[width=0.5\textwidth]{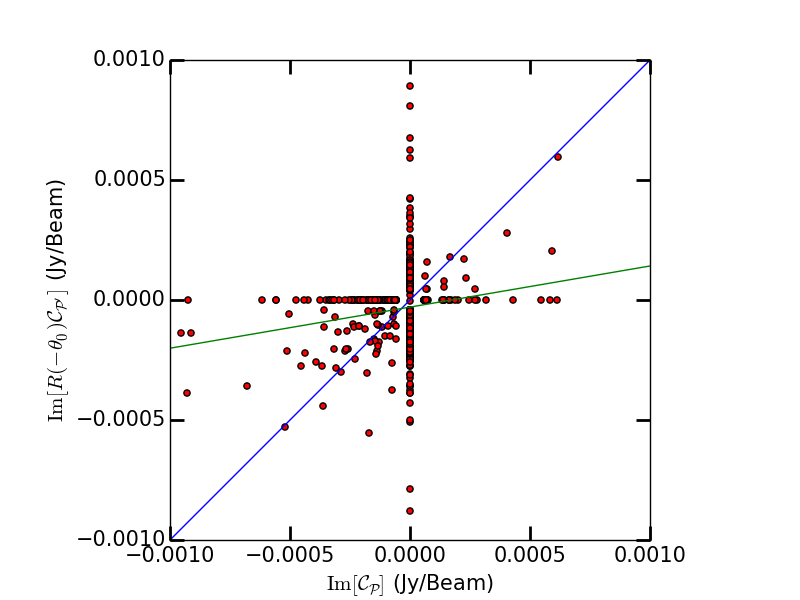}
    \caption{Plots comparing H\"{o}gbom CLEAN components generated in the rotated frame $\mathcal{P}^\prime$ and in the $\mathcal{P}$ frame, for $\mathcal{Q}$ (left) and $\mathcal{U}$ (right). If standard H\"{o}gbom CLEAN method was rotationally invariant, components would lie along the $\mathcal{C}_\mathcal{P}= R(-\theta_0)\mathcal{C}_{\mathcal{P}^\prime}$ relation, shown as the blue line. However, the standard CLEAN method does not follow this relation. The green line represents the least squares straightline fit of the data ${\rm Re}\left[\mathcal{C}_\mathcal{P}\right] = 0.131238478532\times{\rm Re}\left [  R(-\theta_0)\mathcal{C}_{\mathcal{P}^\prime}\right] -3.96884868414 \times 10^{-05}$ for $\mathcal{Q}$ and  ${\rm Im}\left [  R(-\theta_0)\mathcal{C}_{\mathcal{P}^\prime}\right] = 0.171309727716\times{\rm Im}\left [  R(-\theta_0)\mathcal{C}_{\mathcal{P}^\prime}\right] -2.95853041565 \times 10^{-05}$ for $\mathcal{U}$, showing the difference between the ideal fit.}
    \label{fig:hogbomplotCompare}
\end{minipage}
\end{figure*}

\begin{figure*}
\centering
\begin{minipage}[c]{\textwidth}
\centering
    \includegraphics[width=0.5\textwidth]{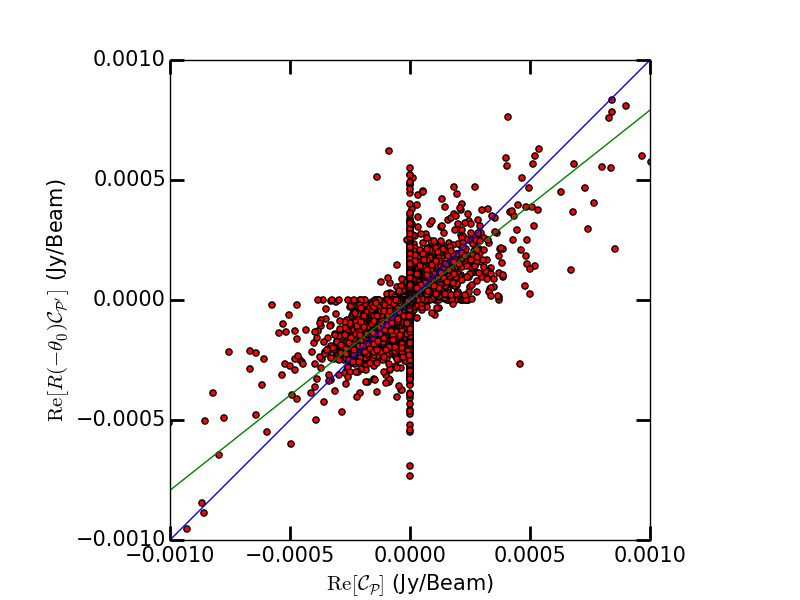}\includegraphics[width=0.5\textwidth]{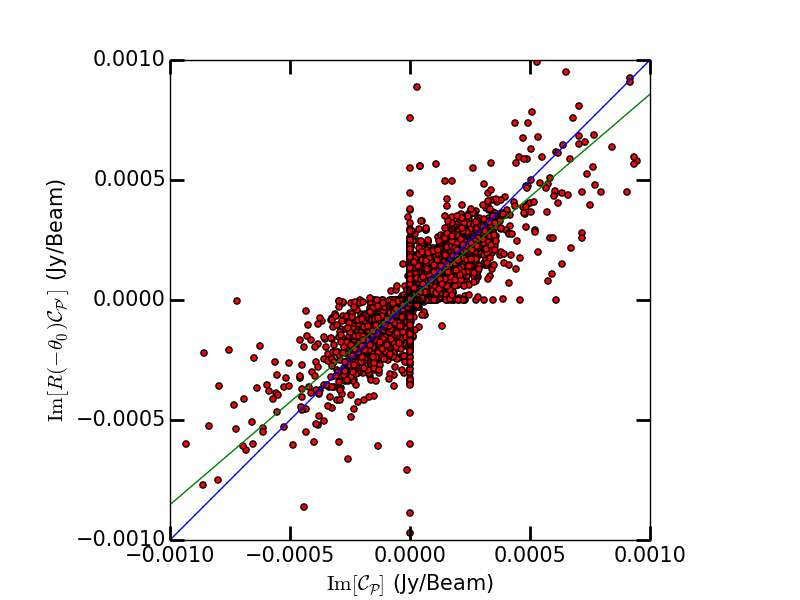}
    \caption{Plots comparing SDI CLEAN components generated in the rotated frame $\mathcal{P}^\prime$ and in the $\mathcal{P}$ frame, for $\mathcal{Q}$ (left) and $\mathcal{U}$ (right). If the complex CLEAN method was rotationally invariant, components would lie along the $\mathcal{C}_\mathcal{P}= R(-\theta_0)\mathcal{C}_{\mathcal{P}^\prime}$ relation, shown as the blue line. While it follows the line more closely than the standard H\"{o}gbom CLEAN method in Figure \ref{fig:hogbomplotCompare}, it still does not follow the line.  The green line represents the least squares straightline fit of the data ${\rm Re}\left[\mathcal{C}_\mathcal{P}\right] = 0.791999348777\times{\rm Re}\left [  R(-\theta_0)\mathcal{C}_{\mathcal{P}^\prime}\right] -8.73557309644 \times 10^{-07} $ for $\mathcal{Q}$ and  ${\rm Im}\left[\mathcal{C}_\mathcal{P}\right] = 0.855227816842\times{\rm Im}\left [  R(-\theta_0)\mathcal{C}_{\mathcal{P}^\prime}\right] + 2.34141194177 \times 10^{-06} $ for $\mathcal{U}$, showing the difference between the ideal fit.}
    \label{fig:steerplotCompare}
\end{minipage}
\end{figure*}

Furthermore, we can more clearly demonstrate the lack of rotational invariance by considering the relation $\mathcal{C}_\mathcal{P} = R(-\theta_0)\mathcal{C}_{\mathcal{P}^\prime}$ and splitting into its real and imaginary components and comparing these for the original and rotated CLEAN processes. Thus in Figures \ref{fig:hogbomplotCompare} \& \ref{fig:steerplotCompare} we show ${\rm Re}\left [ \mathcal{C}_\mathcal{P}\right]$ versus ${\rm Re}\left [  R(-\theta_0)\mathcal{C}_{\mathcal{P}^\prime}\right]$ and ${\rm Im}\left [ \mathcal{C}_\mathcal{P}\right]$ versus ${\rm Im}\left [  R(-\theta_0)\mathcal{C}_{\mathcal{P}^\prime}\right]$, for the H\"{o}gbom and SDI CLEANs, respectively. If these processes were rotationally invariant all of the components would line along the one-to-one line such that ${\rm Re}\left [ \mathcal{C}_\mathcal{P}\right] = {\rm Re}\left [  R(-\theta_0)\mathcal{C}_{\mathcal{P}^\prime}\right]$ and ${\rm Im}\left [ \mathcal{C}_\mathcal{P}\right] = {\rm Im}\left [  R(-\theta_0)\mathcal{C}_{\mathcal{P}^\prime}\right]$, shown in blue on the figures. However we see this is far from the case and components are scattered across the plot. In these examples, the standard H\"{o}gbom CLEAN is slightly more prone to generating biased components than the standard SDI CLEAN, but neither are particularly good.

\section{GENERALISED COMPLEX CLEAN}
\begin{figure*}
\centering
\begin{minipage}[c]{\textwidth}
\centering
    \includegraphics[width=\textwidth]{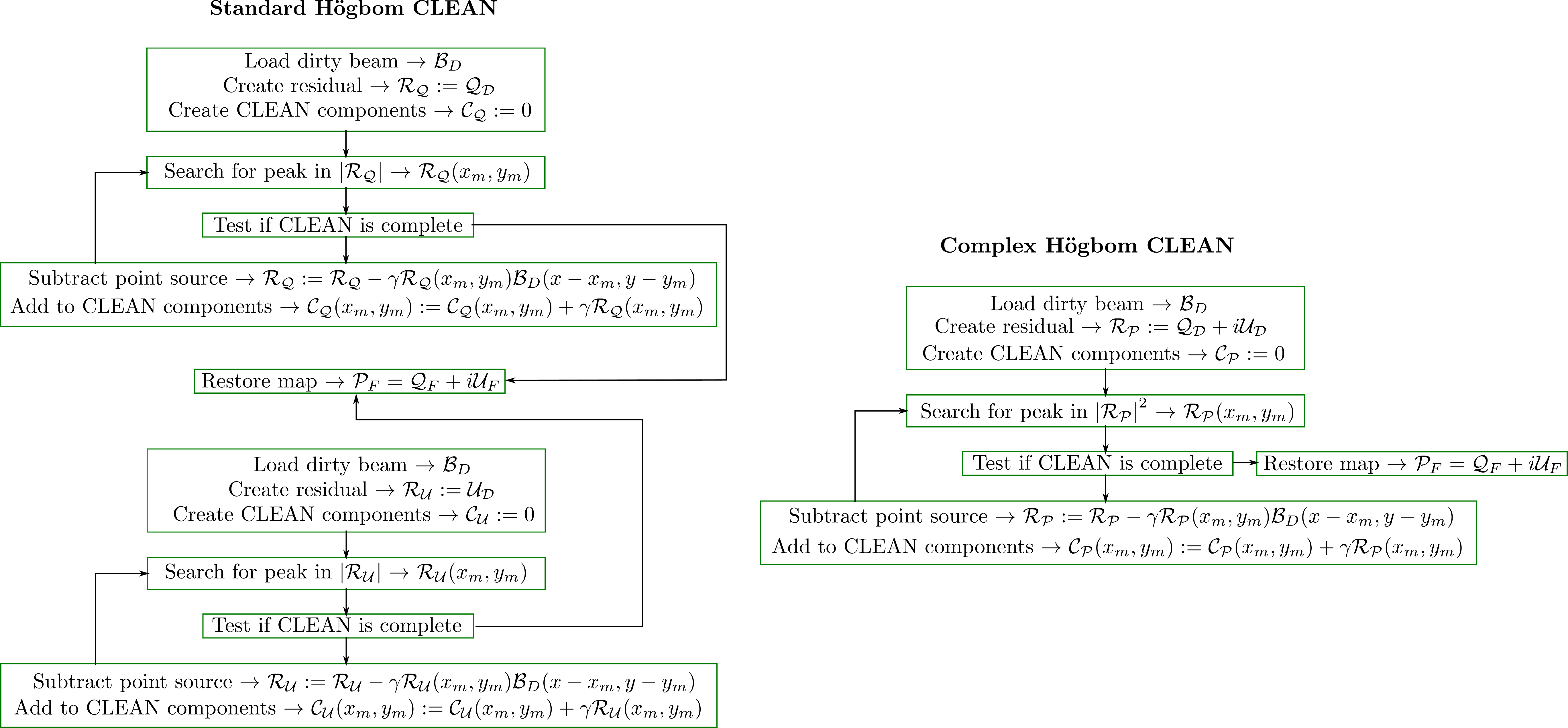}
    \caption{Flow-diagram of the standard H\"{o}gbom CLEAN algorithm, compared to the Complex H\"{o}gbom CLEAN algorithm. The standard H\"{o}gbom CLEAN method deconvolves $\mathcal{Q}$ and $\mathcal{U}$ individually, but the Complex H\"{o}gbom CLEAN deconvolves them together. $\mathcal{B}_D$ is the synthesized beam, $\mathcal{R}$ is the residual image, $\mathcal{C}$ is the CLEAN component image.  $\mathcal{P}_{D,F}$ are the complex linear polarisation synthesized image and final restored image. $\gamma$ is the gain parameter, typically $\gamma:=0.1$.}
    \label{fig:hogbomdiagram}
\end{minipage}
\end{figure*}
\begin{figure*}
\centering
\begin{minipage}[c]{\textwidth}
\centering
    \includegraphics[width=\textwidth]{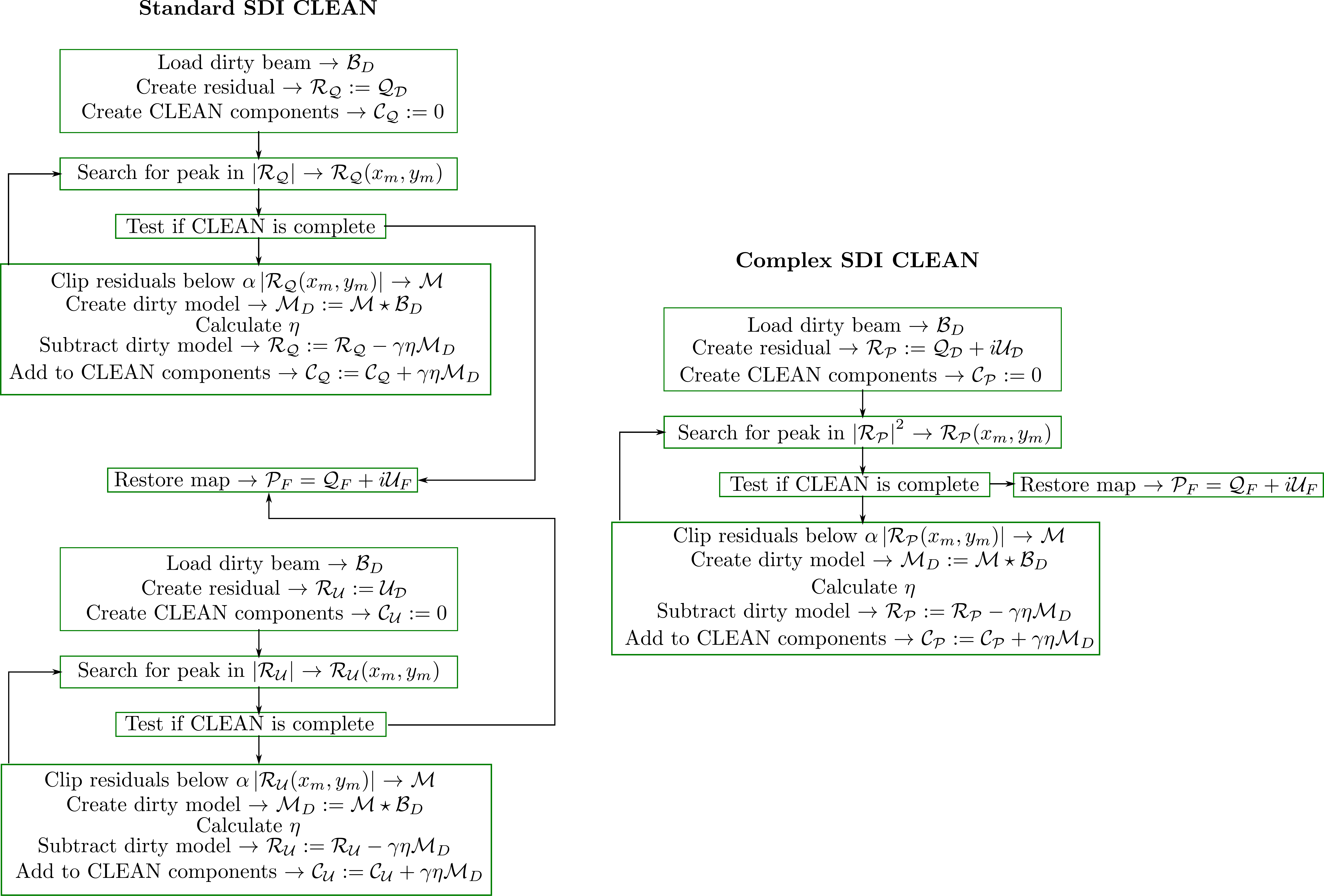}
    \caption{Flow-diagram of the standard SDI CLEAN algorithm, compared to the complex SDI CLEAN algorithm.  The standard SDI CLEAN method deconvolves $\mathcal{Q}$ and $\mathcal{U}$ individually, but the complex SDI CLEAN deconvolves them together. $\mathcal{B}_D$ is the synthesized beam, $\mathcal{R}$ is the residual image, $\mathcal{C}$ is the CLEAN component image. $\mathcal{I}_{D,F}$ are the Stokes $\mathcal{I}$ synthesized image and final restored image. $\mathcal{P}_{D,F}$ are the complex linear polarisation synthesized image and final restored image. $\gamma$ is the gain parameter, typically $\gamma:=0.1$. $\alpha$ is the clip parameter, typically $0.75\leq \alpha \leq 0.99$, and $\mathcal{M}$ is the model of the clipped residual image, and $\mathcal{M}_D$ is the synthesized model. $\eta$ is the scaling of the synthesized model due to the beam volume, where $\eta:=\max\{0.02, \frac{\sum_{k=1}^N \mathcal{R}_\mathcal{P,Q,U}(x_k,y_k)\mathcal{M}_D^\star(x_k,y_k)}{\sum_{k=1}^N \mathcal{M}_D(x_k,y_k)\mathcal{M}_D^\star(x_k,y_k)}\}$.}
    \label{fig:steerdiagram}
\end{minipage}
\end{figure*}
\label{sec:comclean}
In this section we extend the standard CLEAN methods to images of complex linear polarisation. While the methods in this section are simplest to describe using complex numbers, the operations can be translated to operations on the real and imaginary parts in parallel. Where CLEAN assumes that an image is a collection of point sources, the Generalised Complex CLEAN assumes that the image is a collection of point sources with a complex amplitude.

In this work, the synthesized complex polarisation image is defined as
\begin{equation}
\mathcal{P}_{D} =\mathcal{P} \star \mathcal{B}_{D} +\mathcal{N}_\mathcal{Q}+i\mathcal{N}_\mathcal{U} + \mathcal{N}_\mathcal{QU}\, ,
\end{equation}
where $\mathcal{P}$ is the observed complex linear polarisation, $\mathcal{B}_{D}$ is the instruments point spread function (synthesized beam), $\mathcal{P}_{D} $ is the synthesized complex linear polarisation image, and $i=\sqrt{-1}$. $\mathcal{N}_\mathcal{Q}$ and $\mathcal{N}_\mathcal{U}$ are images of Gaussian noise of the same distribution, with variance $\sigma^2$. $\mathcal{N}_\mathcal{QU}$ (complex valued) represents any correlation between the noise in $\mathcal{Q}$ and $\mathcal{U}$, we expect this term to be close to zero after calibration, because the term is typically due to cross talk between linear or circular feeds. Given that linear polarisation is defined in terms of Stokes $\mathcal{Q}$ and $\mathcal{U}$ where $\mathcal{P}=\mathcal{Q}+i\mathcal{U}$, it follows that
\begin{equation}
\mathcal{Q}_{D} =\mathcal{Q} \star \mathcal{B}_{D}+\mathcal{N}_Q + {\rm Re}\{ \mathcal{N}_\mathcal{QU} \}\quad 
\end{equation}
and
\begin{equation}
 \quad \mathcal{U}_{D} =\mathcal{U} \star \mathcal{B}_{D}+\mathcal{N}_U  + {\rm Im}\{ \mathcal{N}_\mathcal{QU} \}\, .
\end{equation}

The objective of the Generalised Complex CLEAN is to recover the observed linear polarisation $\mathcal{P}$ given the synthesized complex linear polarisation image $\mathcal{P}_D$ and instrumental point spread function (synthesized beam) $\mathcal{B}_D$. Following the traditional CLEAN methods, this can be done by generating a clean component image $\mathcal{C}_\mathcal{P}$ such that
\begin{equation}
\mathcal{P}_{D}\approx\mathcal{C}_\mathcal{P}\star \mathcal{B}_{D}\, ,
\end{equation}
where $\mathcal{C}_\mathcal{P}$ is complex valued, and $\mathcal{C}_\mathcal{P}=\mathcal{C}_\mathcal{Q}+i\mathcal{C}_\mathcal{U}$. In the previous section, the standard method to recover $\mathcal{C}_\mathcal{P}$ in this context is to use the standard CLEAN methods on $\mathcal{Q}$ and $\mathcal{U}$ separately to obtain $\mathcal{C}_\mathcal{Q}$ and $\mathcal{C}_\mathcal{U}$ independently. 

The Generalised Complex CLEAN methods start with the residual image equal to the synthesized image, $\mathcal{R}_\mathcal{P}=\mathcal{P}_D$. Each iteration starts by calculating $|\mathcal{R}_\mathcal{P}|^2$. The brightest sources are located in $|\mathcal{R}_\mathcal{P}|^2$, which are then modeled and subtracted from $\mathcal{R}_\mathcal{P}$. The location and peak of each source is recorded as a CLEAN component in $\mathcal{C}_\mathcal{P}$.

When the total number of iterations have been reached, or the residual peak is less than the cutoff $|\mathcal{R}|<\xi$, the CLEAN component image $\mathcal{C}_\mathcal{P}$ is complete and the final residual image $\mathcal{R}_{\mathcal{P}_F}$ has been generated. The restored image is then defined by
\begin{equation}
\mathcal{P}_F = \mathcal{C}_\mathcal{P}\star \mathcal{B}+\mathcal{R}_{\mathcal{P}_F} \, ,
\end{equation}
where $\mathcal{B}$ is a Gaussian beam determined by the resolving power of the observation\footnote{Here we have assumed that the $\mathcal{Q}$ and $\mathcal{U}$ beams are the same. In some situations, such as for significant flagging, or for time switched polarization measurement e.g. quarter wave plates, this may not be true. In these cases there will be an additional bias to spatial scales where the coverage is mismatched. For point sources, the effect is unimportant and more generally speaking, as there is typically not much difference in the $\mathcal{Q}$ and $\mathcal{U}$ beam, the overall effect will be small.}. It is also clear that $\mathcal{Q}$ and $\mathcal{U}$ can be restored separately using the CLEAN components generated from the Generalised Complex CLEAN

\begin{equation}
\mathcal{Q}_F = \mathcal{C}_\mathcal{Q}\star \mathcal{B}+\mathcal{R}_{\mathcal{Q}_F} \quad {\rm and} \quad \mathcal{U}_F = \mathcal{C}_\mathcal{U}\star \mathcal{B}+\mathcal{R}_{\mathcal{U}_F}  \, ,
\end{equation}
where $\mathcal{R}_{\mathcal{Q}_F} = {\rm Re}\left\{\mathcal{R}_{\mathcal{P}_F}\right\}$ and $\mathcal{R}_{\mathcal{U}_F} = {\rm Im}\left\{\mathcal{R}_{\mathcal{P}_F}\right\}$.
Flow-diagrams for the Complex H\"{o}gbom and SDI CLEAN algorithms can be found in Figures \ref{fig:hogbomdiagram} \& \ref{fig:steerdiagram}, where they are compared to their standard counterparts. In the following subsections, we describe the process for a given Complex CLEAN iteration for each of the Complex H\"{o}gbom and Complex SDI CLEANs.

\subsection{Complex H\"{o}gbom CLEAN}
\label{sec:Chogbom}
Each iteration starts by calculating $|\mathcal{R}_\mathcal{P}|^2$, then the position of the peak pixel in $|\mathcal{R}_\mathcal{P}|^2$ is located as $(x_m,y_m)$. If $|\mathcal{R}_\mathcal{P}|^2\leq\xi^2$, or the total number of iterations has been reached, then the Complex CLEAN is complete. Otherwise, the image of the synthesized beam $\mathcal{B}_D$ is translated by $(x_m,y_m)$ and scaled by $\mathcal{R}_\mathcal{P}(x_m,y_m)$. To improve convergence, the synthesized beam is then further scaled by the gain parameter $\gamma$ (typically $\gamma = 0.1$). $\gamma \mathcal{R}_\mathcal{P}(x_m, y_m)$ is added to the CLEAN component $\mathcal{C}_\mathcal{P}(x_m,y_m)$. The translated and scaled synthesized beam $\gamma \mathcal{R}_\mathcal{P}(x_m,y_m)  \mathcal{B}_D (x-x_m,y-y_m)$ is then subtracted from the residual image $\mathcal{R}_\mathcal{P}$.

\subsection{Complex SDI CLEAN}
\label{sec:Csteer}
First, the value of the maximum pixel in $|\mathcal{R}_\mathcal{P}|^2$, $\mathcal{R}_\mathcal{P}(x_m,y_m)$, is measured. If $|\mathcal{R}_\mathcal{P}|^2\leq\xi^2$, or the total number of iterations has been reached, then the Complex CLEAN is complete. A new image, $\mathcal{M}$, is generated by performing a clip on the values in $\mathcal{R}_\mathcal{P}$, using $\alpha$ as the clip parameter, values with a modulus squared below $\alpha^2 |\mathcal{R}_\mathcal{P}(x_m,y_m)|^2$ are set to zero. The clipped image is then convolved with the synthesized beam $\mathcal{M}_D =\mathcal{M}\star \mathcal{B}_D$. 

Because $\mathcal{M}$ will have multiple components, $\mathcal{M}_D$ needs to be scaled by a damping factor (beam volume factor), $\eta$. We choose to calculate $\eta$ following the implementation of SDI CLEAN in MIRIAD, however, adapting the calculation for complex numbers. To estimate $\eta$, the damping factor is calculated by correlating $\mathcal{M}_D$ with $\mathcal{R}_\mathcal{P}$ (a lower limit on $\eta$ may be needed to ensure stability). This correlation is explicitly written as
\begin{equation}
\tilde{\eta} = \frac{\sum_{k=1}^N \mathcal{R}_\mathcal{P}(x_k,y_k)\mathcal{M}_D^\star(x_k,y_k)}{\sum_{k=1}^N \mathcal{M}_D(x_k,y_k)\mathcal{M}_D^\star(x_k,y_k)}\,.
\end{equation}
It is possible to choose $\eta = \tilde{\eta}$. However, $\tilde{\eta}$ may be close to zero and cause a semi-infinite loop of iterations, so we choose to follow MIRIAD by putting a lower limit of magnitude on $\tilde{\eta}$, by defining
\begin{equation}
\eta = {\rm max}\left\{0.02,| \tilde{\eta} |\right\}\frac{\tilde{\eta}}{| \tilde{\eta} |} \, . 
\end{equation}
This choice of calculation is selected so that $\eta$ is consistent when choosing different $\mathcal{Q}$ and $\mathcal{U}$ frames. The lower limit of 0.02 on $|\eta|$ was chosen empirically, and is used in MIRIAD's standard SDI CLEAN method. Using a lower limit of 0.02 may not always be the optimal choice, in which case one may want to choose a different value for the gain $\gamma$.

Then the gain parameter $\gamma$ is used to further scale $\mathcal{M}$ and $\mathcal{M}_D$. $\eta\gamma\mathcal{M}$ is added to the CLEAN component image $\mathcal{C}_\mathcal{P}$, and $\eta\gamma\mathcal{M}_D$ is subtracted from the residual $\mathcal{R}_\mathcal{P}$.

\section{Rotational Invariance of Complex CLEAN}
In this section, we compare the standard and complex CLEAN methods to show that the Generalised Complex CLEAN is invariant under rotations of linear polarisation.  

\begin{figure*}
\centering
\begin{minipage}[c]{\textwidth}
\centering
    \includegraphics[width=0.5\textwidth]{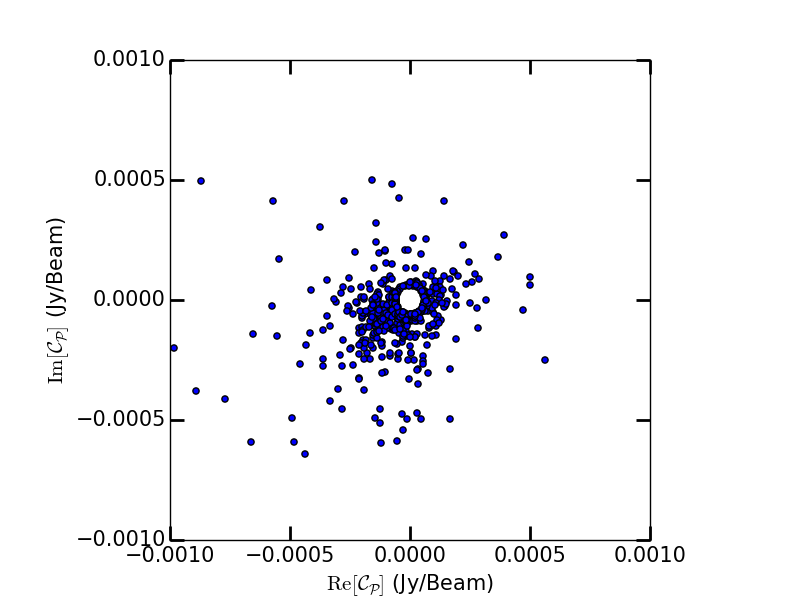}\includegraphics[width=0.5\textwidth]{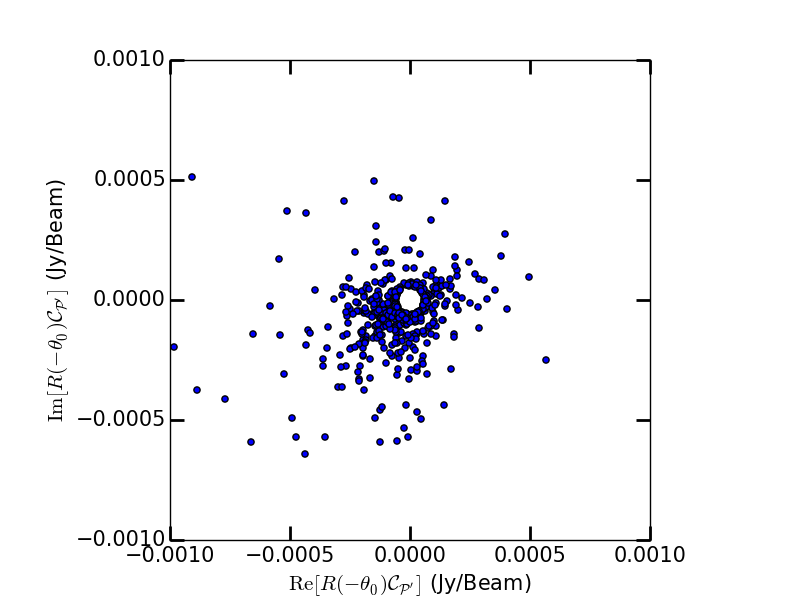}
    \caption{Plots of the linear polarisation states of pixels in the restored source models for PKS J0334-391. The source models were generated from Complex H\"{o}gbom CLEAN components. Left: Polarisation states of the model when the Complex H\"{o}gbom CLEAN components are generated in the $\mathcal{Q}$ and $\mathcal{U}$ frame. Right: Polarisation states of the model when the Complex H\"{o}gbom CLEAN components are generated in the $\mathcal{Q}^\prime$ and $\mathcal{U}^\prime$ frame, where $\mathcal{Q}^\prime$ is the frame rotated by $\theta_0=45^\circ$ from the $\mathcal{Q}$ axis. While there are small variations between the models, there is no preference related to the axes used for deconvolution.}
    \label{fig:Chogbomplot}
\end{minipage}
\end{figure*}

\begin{figure*}
\centering
\begin{minipage}[c]{\textwidth}
\centering
    \includegraphics[width=0.5\textwidth]{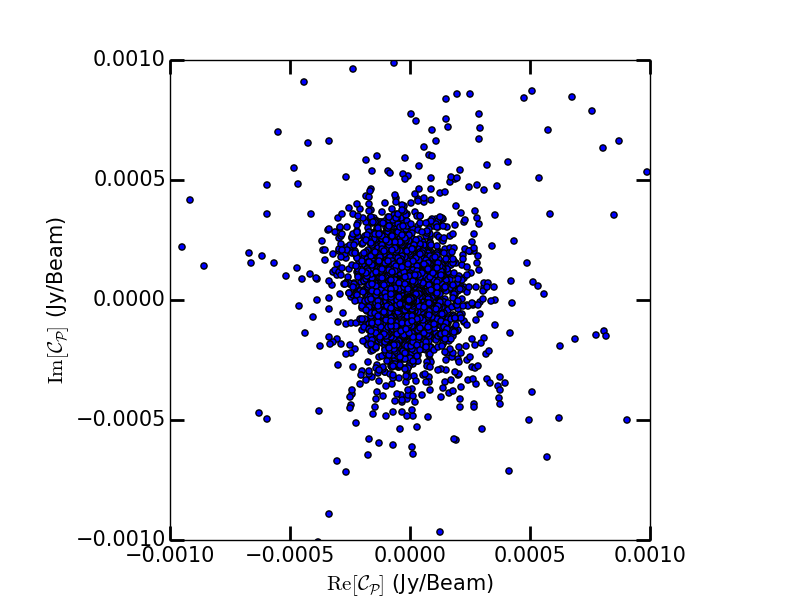}\includegraphics[width=0.5\textwidth]{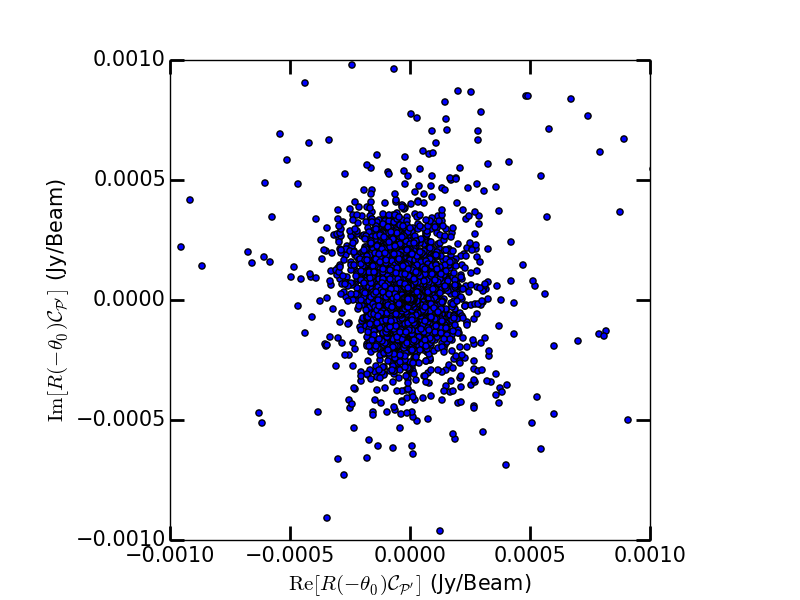}
    \caption{Plots of the linear polarisation states of pixels in the restored source models for PKS B1637-771. The source models were generated from Complex SDI CLEAN components. Left: Polarisation states of the model when the Complex SDI CLEAN components are generated in the $\mathcal{Q}$ and $\mathcal{U}$ frame. Right: Polarisation states of the model when the Complex SDI CLEAN components are generated in the $\mathcal{Q}^\prime$ and $\mathcal{U}^\prime$ frame, where $\mathcal{Q}^\prime$ is the frame rotated by $\theta_0=45^\circ$ from the $\mathcal{Q}$ axis. While there are small variations between the models, there is no preference related to the axes used for deconvolution.}
    \label{fig:Csteerplot}
\end{minipage}
\end{figure*}

As with the standard CLEAN procedures discussed in Section \ref{sec:standard}, we begin by plotting the linear polarisation distribution of the CLEAN components for the original and rotated frames for a Complex H\"{o}gbom CLEAN of the PKS J0334-3900 dataset and a Complex SDI CLEAN of the PKS B1637-771 dataset in Figures \ref{fig:Chogbomplot} and \ref{fig:Csteerplot}, respectively. Again the for the rotated frames (right panels of Figures \ref{fig:Chogbomplot} and \ref{fig:Csteerplot}) the deconvolution process is performed in the $\mathcal{Q}^\prime$ and $\mathcal{U}^\prime$ frame, where $\theta_0 = 45^\circ$ and then after deconvolution, the CLEAN components in the $\mathcal{Q}^\prime$ and $\mathcal{U}^\prime$ frame are rotated back to the original frame. Compared to the standard H\"{o}gbom and SDI plots of the same data in Figures \ref{fig:hogbomplot} \& \ref{fig:steerplot}, here we see firstly that there is no preferred alignment of components with the deconvolution axes and that both the original and rotated plots show almost identical components. This shows that the complex versions are indeed rotationally invariant and thus correcting picking up physically meaningful components regardless of the axes chosen. 

\begin{figure*}
\centering
\begin{minipage}[c]{\textwidth}
\centering
    \includegraphics[width=0.5\textwidth]{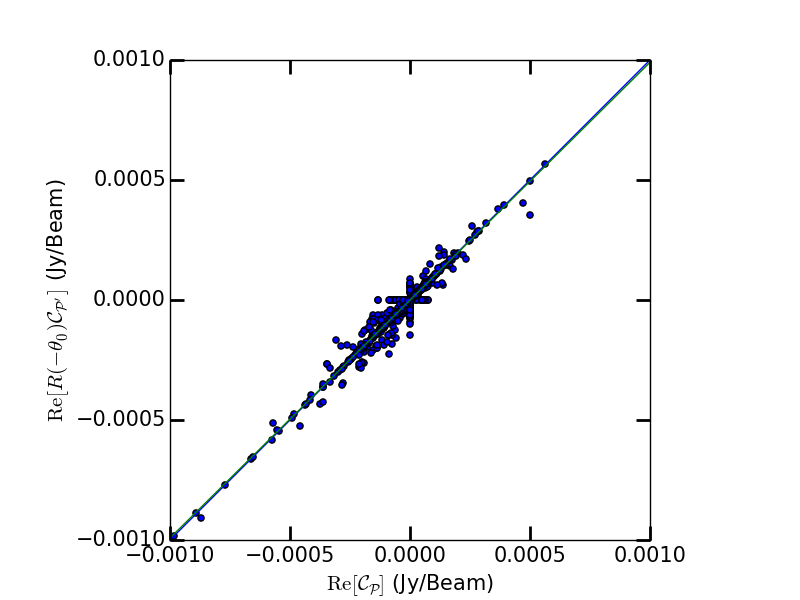}\includegraphics[width=0.5\textwidth]{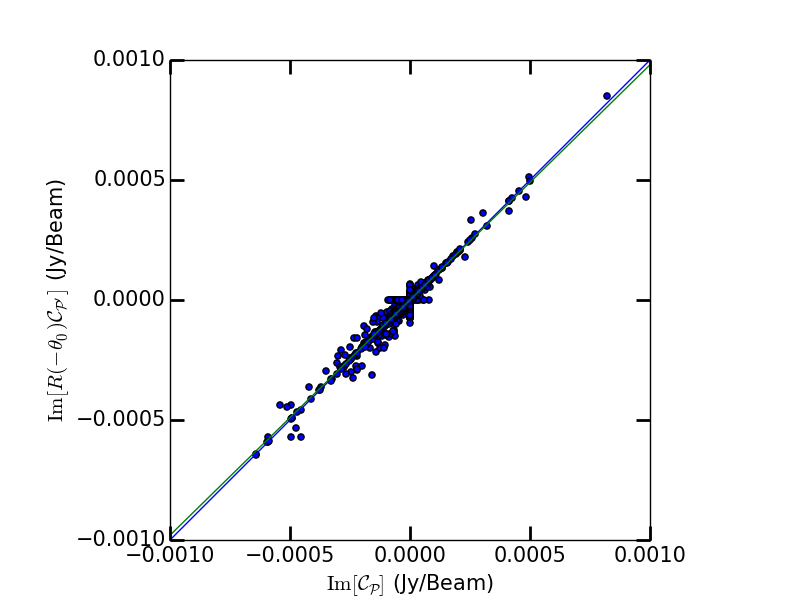}
    \caption{Plots comparing the Complex H\"{o}gbom CLEAN components of PKS J0334-3900 in the $\mathcal{P}^\prime$ and $\mathcal{P}$ frames, for $\mathcal{Q}$ (left) and $\mathcal{U}$ (right). It is expected that the Complex CLEAN method should be rotationally invariant and the components should lie along the $\mathcal{C}_\mathcal{P}= R(-\theta_0)\mathcal{C}_{\mathcal{P}^\prime}$ relation, shown as the blue line. However, deviations from this relation occur as a result of cumulative numerical error in subtractions after many iterations.  The green line represents the least squares straightline fit of the data ${\rm Re}\left[\mathcal{C}_\mathcal{P}\right] = 0.991117836415\times{\rm Re}\left [  R(-\theta_0)\mathcal{C}_{\mathcal{P}^\prime}\right] -7.21750389036 \times 10^{-07}$ for $\mathcal{Q}$ and  ${\rm Im}\left[\mathcal{C}_\mathcal{P}\right] = 0.980077221397\times{\rm Im}\left [  R(-\theta_0)\mathcal{C}_{\mathcal{P}^\prime}\right] -6.47692055102 \times 10^{-08}$ for $\mathcal{U}$, showing the difference between the ideal fit.}
    \label{fig:ChogbomplotCompare}
\end{minipage}
\end{figure*}

\begin{figure*}
\centering
\begin{minipage}[c]{\textwidth}
\centering
    \includegraphics[width=0.5\textwidth]{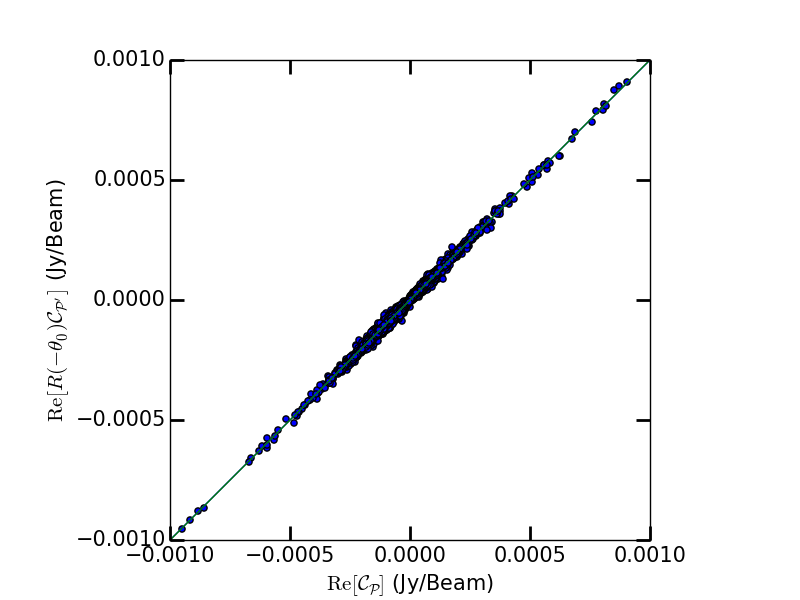}\includegraphics[width=0.5\textwidth]{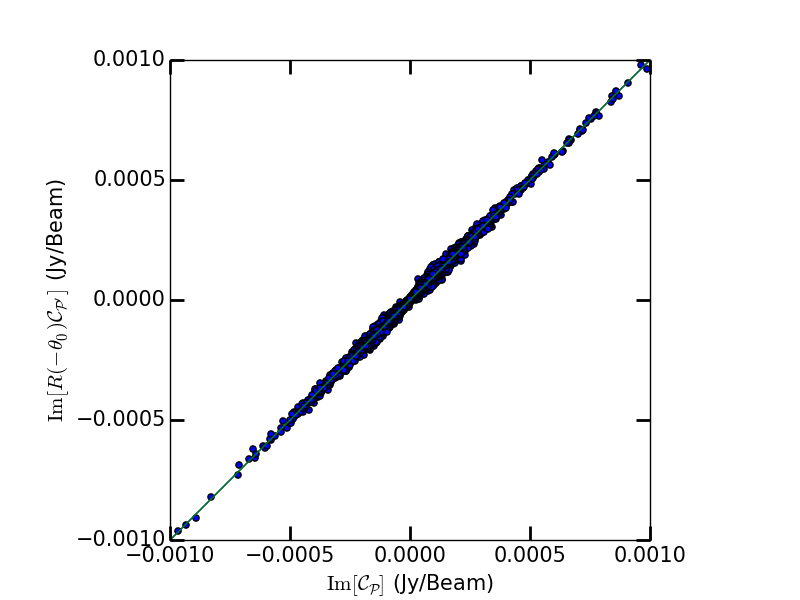}
    \caption{Plots comparing the Complex SDI CLEAN components of PKS B1637-771 generated in the $\mathcal{P}^\prime$ and $\mathcal{P}$ frames, for $\mathcal{Q}$ (left) and $\mathcal{U}$ (right). It is expected that the complex CLEAN method should be rotationally invariant and the components should lie along the $\mathcal{C}_\mathcal{P}= R(-\theta_0)\mathcal{C}_{\mathcal{P}^\prime}$ relation, shown as the blue line. It is known that the SDI CLEAN is less prone to cumulative numerical error when compared to the H\"{o}gbom CLEAN, because it subtracts many components in a single iteration. This is demonstrated by the components following the $\mathcal{C}_\mathcal{P}= R(-\theta_0)\mathcal{C}_{\mathcal{P}^\prime}$ relation more closely than in Figure \ref{fig:ChogbomplotCompare}. The green line represents the least squares straightline fit of the data ${\rm Re}\left[\mathcal{C}_\mathcal{P}\right] = 1.00046246231\times{\rm Re}\left [  R(-\theta_0)\mathcal{C}_{\mathcal{P}^\prime}\right] -2.84010989593 \times 10^{-08}$ for $\mathcal{Q}$ and  ${\rm Re}\left[\mathcal{C}_\mathcal{P}\right] = 1.00059996577\times{\rm Im}\left [  R(-\theta_0)\mathcal{C}_{\mathcal{P}^\prime}\right] -2.24977985489 \times 10^{-08}$ for $\mathcal{U}$, showing the difference between the ideal fit.}
    \label{fig:CsteerplotCompare}
\end{minipage}
\end{figure*}

As before we also demonstrate the rotational invariance by considering the relation $\mathcal{C}_\mathcal{P} = R(-\theta_0)\mathcal{C}_{\mathcal{P}^\prime}$ and splitting into its real and imaginary components and comparing these for the original and rotated Complex CLEAN processes. Figures \ref{fig:ChogbomplotCompare} \& \ref{fig:CsteerplotCompare} show ${\rm Re}\left [ \mathcal{C}_\mathcal{P}\right]$ versus ${\rm Re}\left [  R(-\theta_0)\mathcal{C}_{\mathcal{P}^\prime}\right]$ and ${\rm Im}\left [ \mathcal{C}_\mathcal{P}\right]$ versus ${\rm Im}\left [  R(-\theta_0)\mathcal{C}_{\mathcal{P}^\prime}\right]$, for the Complex H\"{o}gbom and Complex SDI CLEANs, respectively. Unlike Figures \ref{fig:hogbomplotCompare} \& \ref{fig:steerplotCompare} which use the same data cleaned with the standard methods, here we see the Complex CLEAN components line along the one-to-one line such that ${\rm Re}\left [ \mathcal{C}_\mathcal{P}\right] = {\rm Re}\left [  R(-\theta_0)\mathcal{C}_{\mathcal{P}^\prime}\right]$ and ${\rm Im}\left [ \mathcal{C}_\mathcal{P}\right] = {\rm Im}\left [  R(-\theta_0)\mathcal{C}_{\mathcal{P}^\prime}\right]$, shown in blue on the Figures. We plot the least squares fit to the data on both plots in green and find it to be in excellent agreement with the one-to-one relation in the case of the Complex H\"{o}gbom CLEAN and identical for the Complex SDI CLEAN. Again, this demonstrates that the Complex CLEAN methods are rotationally invariant.

\begin{figure*}
\centering
\begin{minipage}[c]{\textwidth}
\centering
    \includegraphics[width=0.5\textwidth]{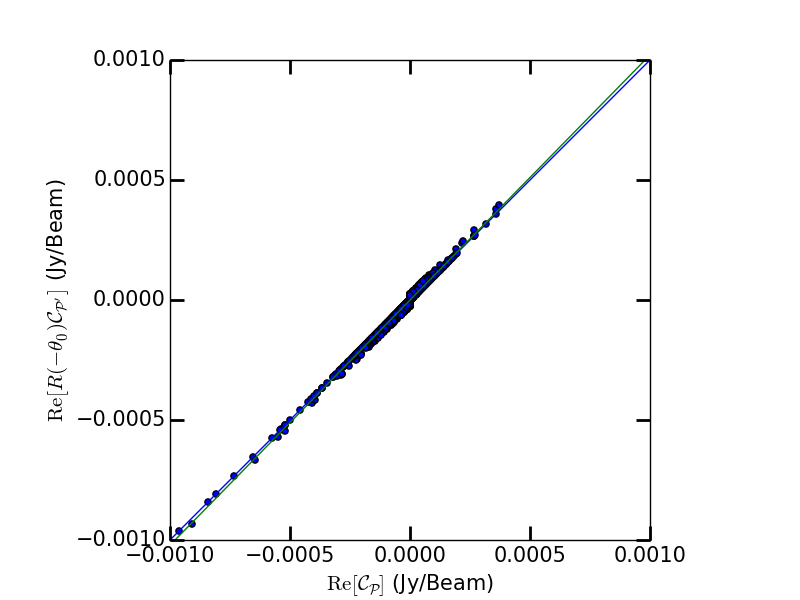}\includegraphics[width=0.5\textwidth]{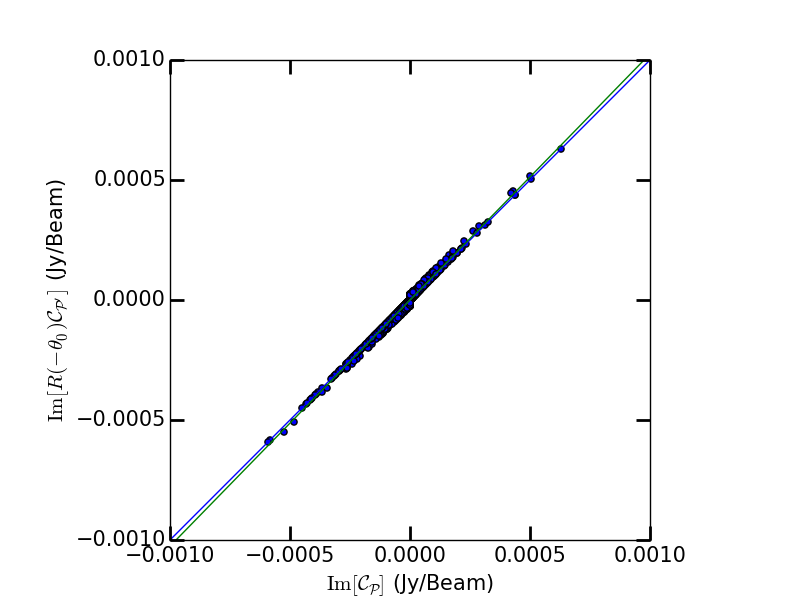}
    \caption{Plots comparing Complex SDI CLEAN components of PKS J0334-3900 generated in the $\mathcal{P}^\prime$ and $\mathcal{P}$ frames, for $\mathcal{Q}$ (left) and $\mathcal{U}$ (right). The blue line is the ideal relation $\mathcal{C}_\mathcal{P}= R(-\theta_0)\mathcal{C}_{\mathcal{P}^\prime}$ and the green line represents the least squares straight line fit of the data. It is clear that the Complex SDI CLEAN method has less scatter than the Complex H\"{o}gbom CLEAN method in Figure \ref{fig:ChogbomplotCompare}, since it more closely follows the ideal relation. The green line follows the least squares fit ${\rm Re}\left[\mathcal{C}_\mathcal{P}\right] = 1.0212280228\times{\rm Re}\left [  R(-\theta_0)\mathcal{C}_{\mathcal{P}^\prime}\right] -2.26837377937 \times 10^{-07}$ for $\mathcal{Q}$ and ${\rm Im}\left[\mathcal{C}_\mathcal{P}\right] = 1.02559965762\times{\rm Im}\left [  R(-\theta_0)\mathcal{C}_{\mathcal{P}^\prime}\right] -1.99039417418 \times 10^{-07}$ for $\mathcal{U}$.}
    \label{fig:ChogbomplotsteerCompare}
\end{minipage}
\end{figure*}

Here the scatter from the ideal case starts to become physically meaningful as a method for characterisation of numerical error in the CLEANing process. From the plots it is clear that the Complex SDI CLEAN has less scatter than the Complex H\"{o}gbom CLEAN and this is consistent with the fact that an SDI CLEAN will produce less numerical error \citep{ste84} after performing a large number of iterations, an advantage over the H\"{o}gbom CLEAN.  To further test this we also performed a Complex SDI clean of PKS J0334-3900 and provide the ${\rm Re}\left [ \mathcal{C}_\mathcal{P}\right]$ versus ${\rm Re}\left [  R(-\theta_0)\mathcal{C}_{\mathcal{P}^\prime}\right]$ and ${\rm Im}\left [ \mathcal{C}_\mathcal{P}\right]$ versus ${\rm Im}\left [  R(-\theta_0)\mathcal{C}_{\mathcal{P}^\prime}\right]$ plots in Figure \ref{fig:ChogbomplotsteerCompare}. Comparing Figures \ref{fig:ChogbomplotCompare} and \ref{fig:ChogbomplotsteerCompare} shows a stark difference in the scatter which is due to increasing numerical errors associated with underlying single component iterative approach of the H\"{o}gbom CLEAN over the multiple component cycles used in a SDI clean. Numerically we find that the sample standard deviation (scatter) about the one-to-one line is of order 130$\mu$Jy for both the real and imaginary parts using the standard H\"{o}gbom CLEAN of PKS J0334-3900. This improves to around 14$\mu$Jy when performing a Complex H\"{o}gbom CLEAN of PKS J0334-3900 and improves further still to around 5$\mu$Jy for a Complex SDI CLEAN on the same data. We see a similar result for the more complex source, PKS B1637-771 which improves from $\sim$40$\mu$Jy to $\sim$4$\mu$Jy in comparing a standard to a Complex SDI CLEAN. Given that this is measuring scatter in the CLEAN components, this is demonstrating the improvement in the accuracy of the magnitude of the components considered. 

\begin{figure*}
\centering
\begin{minipage}[c]{\textwidth}
\centering
    \includegraphics[width=0.5\textwidth]{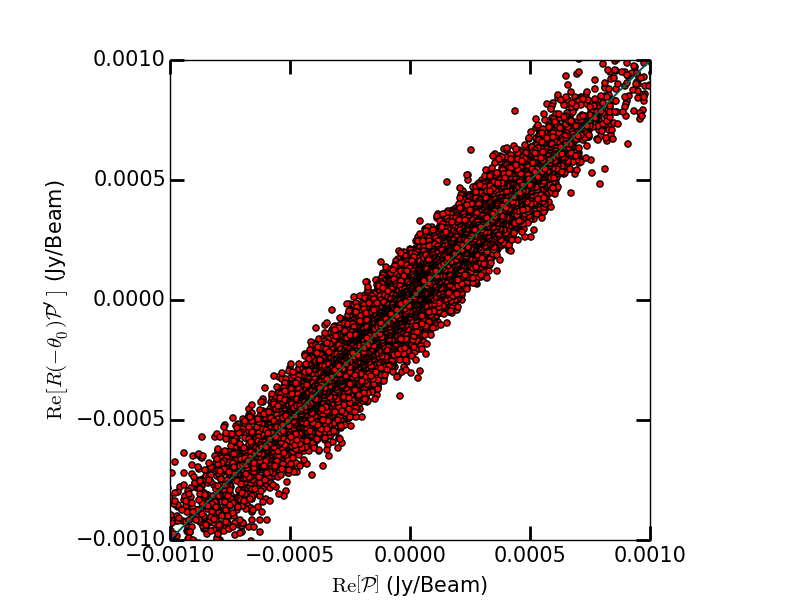}\includegraphics[width=0.5\textwidth]{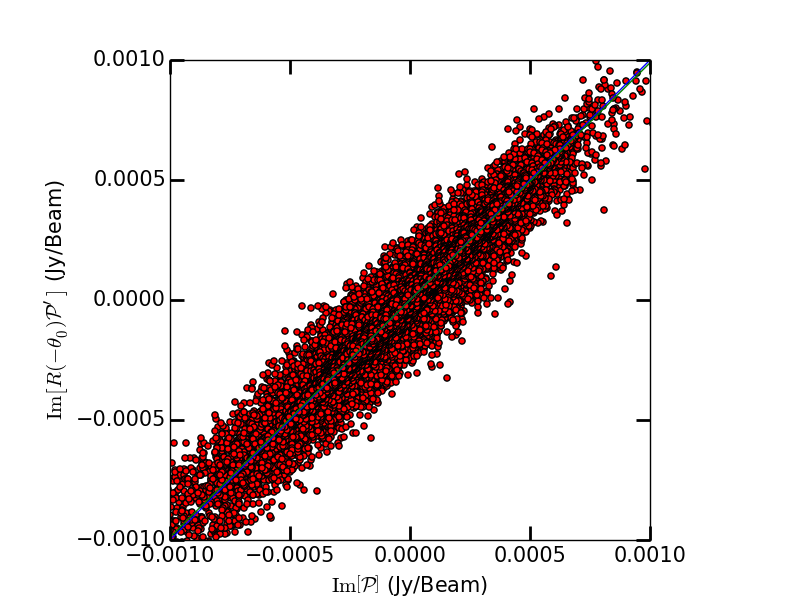}
    \caption{Plots comparing restored images generated from H\"{o}gbom CLEAN components of PKS J0334-3900 in the rotated frame $\mathcal{P}^\prime$ and in the $\mathcal{P}$ frame, for $\mathcal{Q}$ (left) and $\mathcal{U}$ (right). If standard H\"{o}gbom CLEAN method was rotationally invariant, all the data points would lie along the $\mathcal{P}= R(-\theta_0){\mathcal{P}^\prime}$ relation, shown as the blue line. It is clear that the data points are scattered along this line.}
    \label{fig:hogbomplotCompareMag}
\end{minipage}
\end{figure*}

\begin{figure*}
\centering
\begin{minipage}[c]{\textwidth}
\centering
    \includegraphics[width=0.5\textwidth]{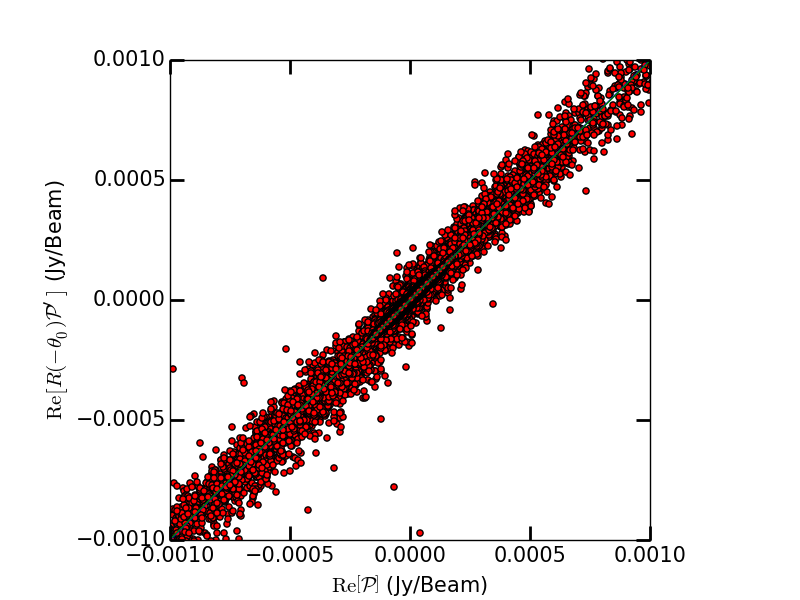}\includegraphics[width=0.5\textwidth]{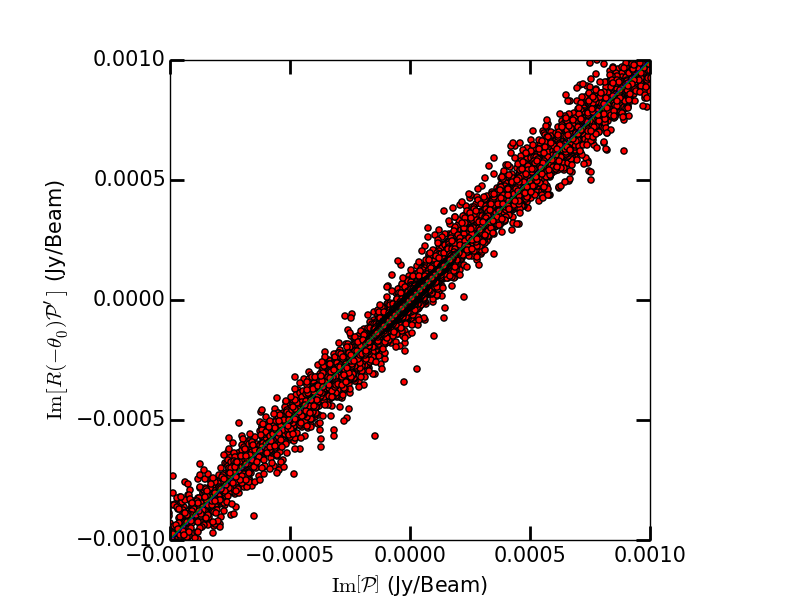}
    \caption{Plots comparing restored images generated from SDI CLEAN components of PKS B1637-771 in the rotated frame $\mathcal{P}^\prime$ and in the $\mathcal{P}$ frame, for $\mathcal{Q}$ (left) and $\mathcal{U}$ (right). If the complex CLEAN method was rotationally invariant, all points would lie along the $\mathcal{P}= R(-\theta_0){\mathcal{P}^\prime}$ relation, shown as the blue line. It is clear that there is less scatter than the standard H\"{o}gbom CLEAN method in Figure \ref{fig:hogbomplotCompareMag}, but the points still do not follow the line closely.  }
    \label{fig:steerplotCompareMag}
\end{minipage}
\end{figure*}

\begin{figure*}
\centering
\begin{minipage}[c]{\textwidth}
\centering
    \includegraphics[width=0.5\textwidth]{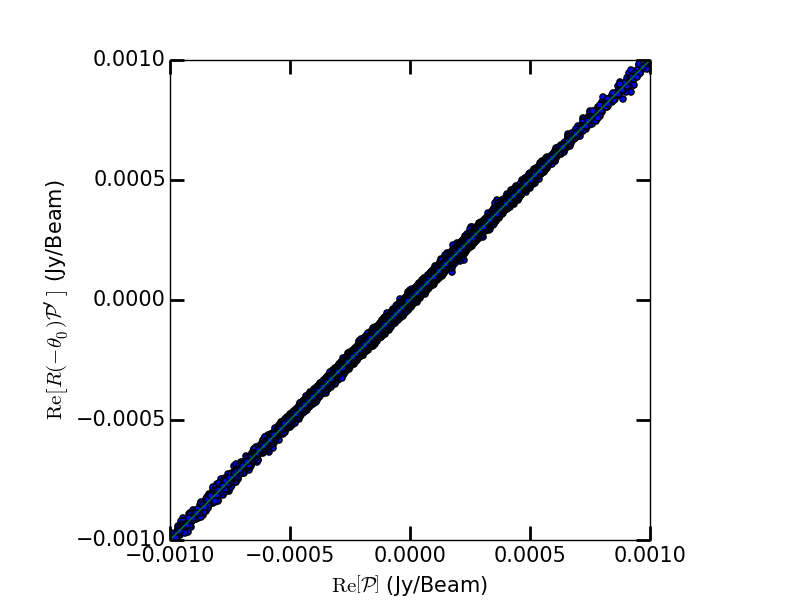}\includegraphics[width=0.5\textwidth]{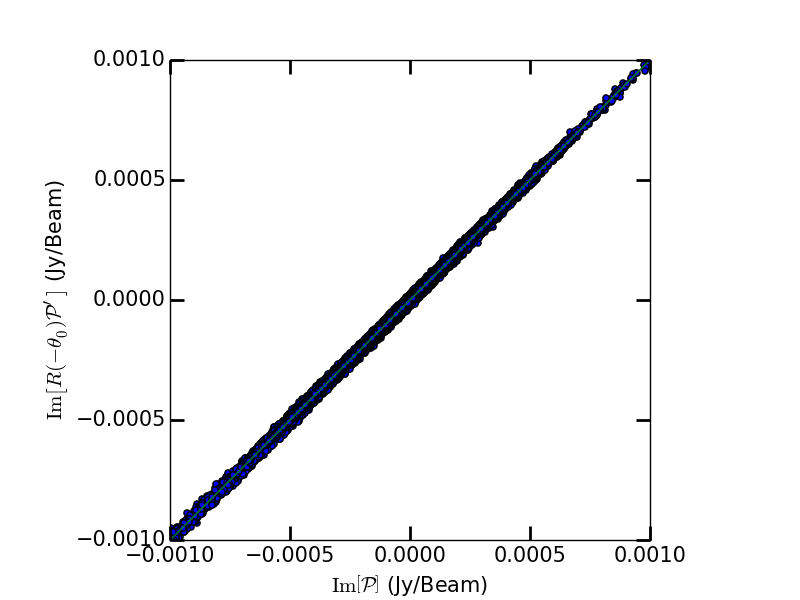}
    \caption{Plots comparing images restored from Complex H\"{o}gbom CLEAN components of PKS J0334-3900 generated in the $\mathcal{P}^\prime$ and $\mathcal{P}$ frames, for $\mathcal{Q}$ (left) and $\mathcal{U}$ (right). It is expected that the complex CLEAN method should be rotationally invariant, and should lie along the $\mathcal{P}= R(-\theta_0){\mathcal{P}^\prime}$ relation, shown as the blue line. It is clear that the Complex H\"{o}gbom CLEAN has less scatter than the non-complex method.}
    \label{fig:ChogbomplotCompareMag}
\end{minipage}
\end{figure*}

\begin{figure*}
\centering
\begin{minipage}[c]{\textwidth}
\centering
    \includegraphics[width=0.5\textwidth]{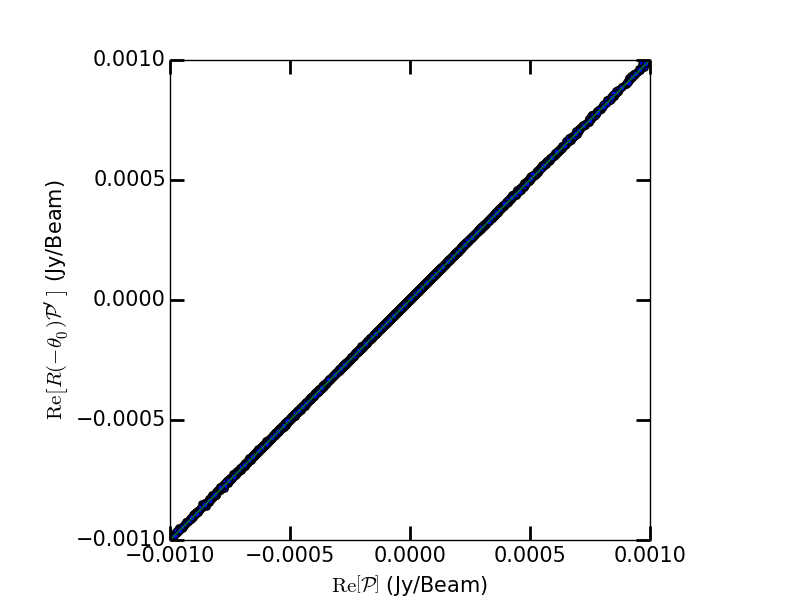}\includegraphics[width=0.5\textwidth]{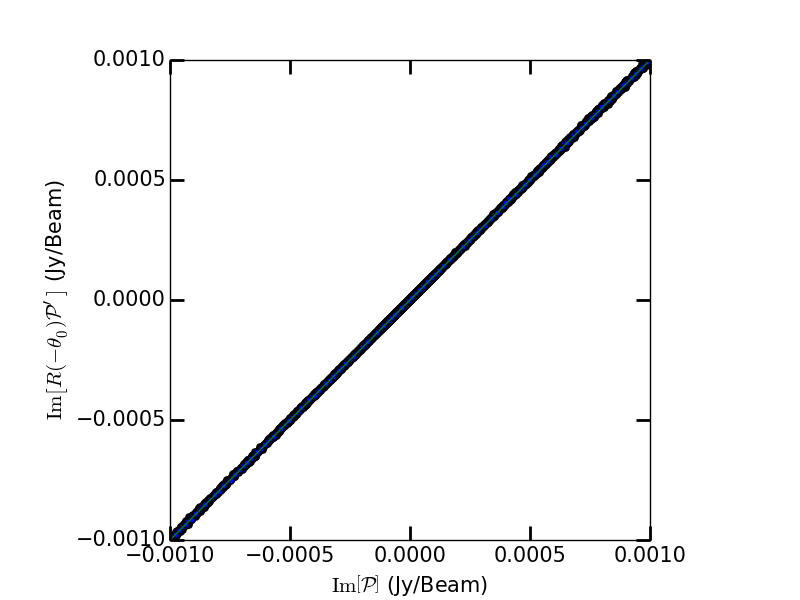}
    \caption{Plots comparing images restored from complex SDI CLEAN components of PKS B1637-771 generated in the $\mathcal{P}^\prime$ and $\mathcal{P}$ frames, for $\mathcal{Q}$ (left) and $\mathcal{U}$ (right). It is expected that the complex CLEAN method should be rotationally invariant, and should lie along the $\mathcal{P}= R(-\theta_0){\mathcal{P}^\prime}$ relation, shown as the blue line. It is clear that the Complex SDI CLEAN has less scatter than the non-complex method and less scatter than the Complex H\"{o}gbom CLEAN. }
    \label{fig:CsteerplotCompareMag}
\end{minipage}
\end{figure*}

In order to illustrate the effect of rotational invariance on the pixel values in the final images of $\mathcal{Q}$ (Re[$\mathcal{P}_F$]) and $\mathcal{U}$ (Re[$\mathcal{P}_F$]) rather than the CLEAN components, we plot the pixel values in the resultant images in the original and rotated frames against each other. Here we do this for a standard H\"{o}gbom CLEAN of PKS J0334-3900 (Figure \ref{fig:hogbomplotCompareMag}), a Complex H\"{o}gbom CLEAN of PKS J0334-3900 (Figure \ref{fig:steerplotCompareMag}), a standard SDI CLEAN of PKS B1637-771 (Figure \ref{fig:ChogbomplotCompareMag}), and the Complex SDI CLEAN of PKS B1637-771 (Figure \ref{fig:CsteerplotCompareMag}). 

As with the CLEAN component comparison we see a very large scatter about the one-to-one relation for pixel values in both the standard H\"{o}gbom and SDI CLEANs with the H\"{o}gbom CLEAN showing the greatest change in pixel values simply due to rotation of the frame in which the CLEAN process is performed. With the Complex H\"{o}gbom and SDI cleans there is considerable improvement and the SDI pixel values are identical between the rotated and non-rotated frames, while the H\"{o}gbom CLEAN values show some scatter which is again due to the increased numerical errors of the method.

\section{Advantages in the Selection of Cut-off values for Components in Complex CLEAN}
\label{sec:cutoff}
Having shown that the Generalised Complex CLEAN methods are rotationally invariant, here we discuss the implications of the rotational invariance to application of cut-off values for CLEAN components in the Generalised Complex CLEAN methods and how this differs from the current methods.

The standard CLEAN methods are used to deconvolve images of complex linear polarisation, by CLEANing and restoring the real and imaginary images separately. Unlike Stokes $\mathcal{I}$, Stokes $\mathcal{Q}$ and $\mathcal{U}$, are expected to have sources of positive and negative values, so the standard CLEAN methods are required to locate peaks of the absolute maximum. In this case, we choose the cutoff to be $\xi := 3\sigma$, where $\sigma$ is the standard deviation of the noise in Stokes $\mathcal{Q}$ and $\mathcal{U}$, which is typically the same value. Both Stokes $\mathcal{Q}$ and $\mathcal{U}$ are cleaned until $\left|\mathcal{R}_{\mathcal{Q}}  \right| < 3\sigma$ and $\left|\mathcal{R}_{\mathcal{U}} \right|<3\sigma$ in each residual image. Furthermore, it is possible for CLEAN to detect a different numbers of components in $\mathcal{C}_\mathcal{Q}$ and $\mathcal{C}_\mathcal{U}$, since the component images are generated separately.

As described in Section \ref{sec:comclean}, the Generalised Complex CLEAN methods will generate the component images $\mathcal{C}_\mathcal{Q}$ and $\mathcal{C}_\mathcal{U}$ simultaneously, and each image will have the same amount of CLEANed components. Since the CLEAN components are located in the peaks of $|\mathcal{R}|$, the Generalised Complex CLEAN will run until $|\mathcal{R}_{\mathcal{Q}}+i\mathcal{R}_{\mathcal{U}}|<3\sigma$.

The advantage of the Generalised Complex CLEAN is that the deconvolution method is rotationally invariant in $(\mathcal{Q},\mathcal{U})$ meaning that in each iteration, the peaks are located in $|\mathcal{R}_{\mathcal{Q}}+i\mathcal{R}_{\mathcal{U}}|$ which is also rotationally invariant under rotations $R(\theta_0$). Thus, the peak amplitude and location for each iteration will also be invariant under rotations and furthermore the cutoff requirement for the Generalised Complex CLEAN is also rotationally invariant, since $\sigma$ is the same in the $\mathcal{Q}$ and $\mathcal{U}$.

\begin{figure}
\centering
\begin{minipage}[c]{0.5\textwidth}
\centering
    \includegraphics[width=\textwidth]{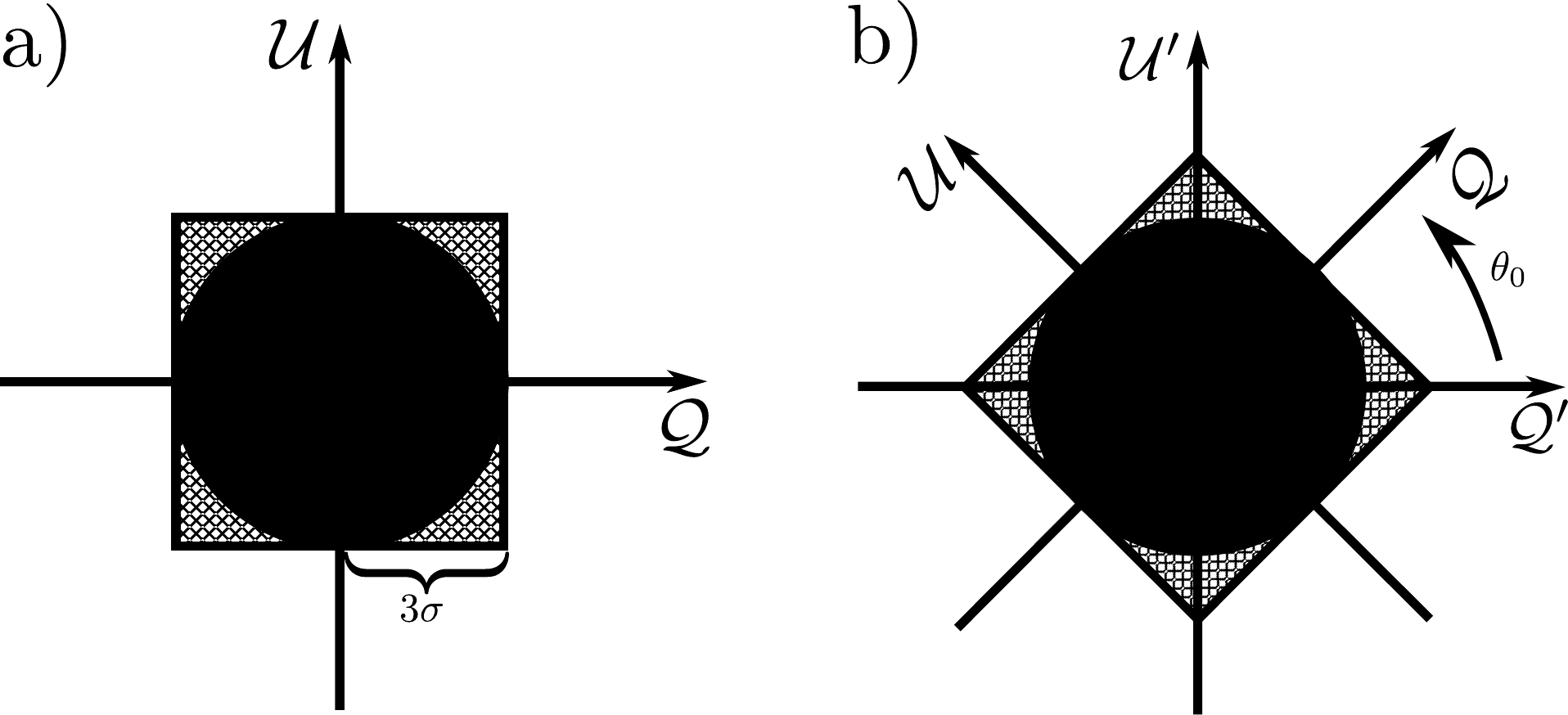}
    \caption{a) and b) show the phase space of ($\mathcal{Q}, \mathcal{U}$). a) The black circle of radius $3\sigma$ represents polarisation states below the detection threshold of $|\mathcal{P}(x,y)|>3\sigma$. The surface of the circle represents the cutoff for subtracting complex CLEAN components. The square represents polarisation states at the detection level of 3$\sigma$ in $\mathcal{Q}$ or $\mathcal{U}$. The polarisation states outside the square are CLEANed using the standard CLEAN method, while the thatched region and the region outside the square are CLEANed using the complex CLEAN method. b) When we rotate the ($\mathcal{Q}, \mathcal{U}$) frame by an angle $\theta_0$ to view the constraints in the $(\mathcal{Q}^\prime, \mathcal{U}^\prime)$ frame it is clear that the standard CLEAN cutoff is not the same for ($\mathcal{Q}, \mathcal{U}$) as for $(\mathcal{Q}^\prime, \mathcal{U}^\prime)$. }
    \label{fig:constraintDiagram}
\end{minipage}
\end{figure}

Additionally, if a component satisfies $|\mathcal{P}(x,y)|>3\sigma$, then one can always rotate to a $(\mathcal{Q}^\prime,\mathcal{U}^\prime)$ frame where $|\mathcal{Q}^\prime (x,y)|>3\sigma$. Therefore, when $|\mathcal{R}_{\mathcal{Q}}+i\mathcal{R}_{\mathcal{U}}|<3\sigma$ is chosen as a cutoff requirement, the Generalised Complex CLEAN subtracts components in $\mathcal{Q}$ and $\mathcal{U}$ above $3\sigma$ in all possible $(\mathcal{Q},\mathcal{U})$ frames. By comparison the standard CLEAN methods have a cut-off that does depend on the choice of $\theta_0$. Specifically, the standard CLEAN method will not detect components when both
\begin{equation}
\frac{3}{\sqrt{2}}\sigma < |\mathcal{Q}|\leq 3\sigma \quad {\rm and} \quad \frac{3}{\sqrt{2}}\sigma < |\mathcal{U}|\leq 3\sigma\, .
\end{equation}
However, these components are detected by the Generalised Complex CLEAN. Figure \ref{fig:constraintDiagram} shows these constrains against states of possible linear polarisation, where the Generalised Complex CLEAN cut-off is compared to the standard CLEAN cut-off. The effect of this is that when using a standard CLEAN, in order to reach components which are significant (e.g. $|\mathcal{P}(x,y)|>3\sigma$) but in regions which are unreachable with a 3$\sigma$ cut off in $\mathcal{Q}$ and $\mathcal{U}$, the cut-off is typically relaxed by increasing the number of iterations and hence components. However, this also results in collecting a very large number of spurious CLEAN components which have $|\mathcal{P}(x,y)|<3\sigma$. In practice one will not actually be able to collect all of the components for which $|\mathcal{P}(x,y)|>3\sigma$ without having an unworkably large number of spurious components and so typically something in the middle is reached were some significant components are missing and a large number of spurious components are included. This scenario both increases the computational requirements and reduces the quality of the final image as compared to a Generalised Complex CLEAN.

\subsection{Cut-off in the case of $\mathcal{Q}$$\mathcal{U}$ correlations} % (fold)
\label{sub:clean_cut_off_in_the_case_of_}

When there are correlations between Stokes $\mathcal{Q}$ and $\mathcal{U}$, which can be caused by instrumental polarisation, the Generalised Complex CLEAN cut-off will change as a function of polariastion angle. While the cut-off is determined by the RMS noise of the signal in the image, it is the variance of the noise that is linear under addition and subtraction of signals. For this reason, the covariance matrix is used to determine how the RMS noise changes under rotations of $\mathcal{P}$. We can write the non-diagonal covariance matrix as
\begin{equation}
\Sigma=\begin{bmatrix}\sigma_{\mathcal{Q}}^2 & \sigma_{\mathcal{Q}\mathcal{U}}^2 \\ \sigma_{\mathcal{Q}\mathcal{U}}^2 & \sigma_{\mathcal{U}}^2 \end{bmatrix}\, .
\end{equation}
Physically, the eigenvectors of the covariance matrix can be used to determine the direction of the instrumental polarisation, assuming the instrumental polarisation causes the signal's polarisation to lie along a preferential direction.

We can change the basis of the covariance matrix to see how the coefficients vary as a function of angle

\begin{widetext}
\begin{equation}
\Sigma^\prime(\theta)=\begin{bmatrix}\sigma_{\mathcal{Q}}^2\cos^2\theta + \sigma_{\mathcal{U}}^2\sin^2\theta + \sigma_{\mathcal{U}\mathcal{Q}}^2\sin 2\theta & \sigma_{\mathcal{Q}\mathcal{U}}^2\cos 2\theta + \frac{\sigma_{\mathcal{Q}}^2 - \sigma_{\mathcal{U}}^2}{2}\sin 2\theta \\ \sigma_{\mathcal{Q}\mathcal{U}}^2\cos 2\theta + \frac{\sigma_{\mathcal{Q}}^2 - \sigma_{\mathcal{U}}^2}{2}\sin 2\theta & \sigma_{\mathcal{U}}^2\cos^2\theta + \sigma_{\mathcal{Q}}^2\sin^2\theta - \sigma_{\mathcal{U}\mathcal{Q}}^2\sin 2\theta \end{bmatrix}\, .
\end{equation}
\end{widetext}

For any given component at angle $\theta_c$, we can rotate our basis so that it lies along the new $\mathcal{Q}^\prime$ axis. In this frame, the RMS noise in $\mathcal{Q}^\prime$ is

\begin{equation}
	\sigma_{\mathcal{Q}^\prime} = \sqrt{\sigma_{\mathcal{Q}}^2\cos^2\theta_c + \sigma_{\mathcal{U}}^2\sin^2\theta_c + \sigma_{\mathcal{U}\mathcal{Q}}^2\sin 2\theta_c}\, ,
\end{equation}
 which is used to determine if the component is above three times the RMS noise level.
Therefore, the cut-off as a function of the component's polarisation angle is
\begin{equation}
	\xi(\theta_c) = 3\sqrt{\sigma_{\mathcal{Q}}^2\cos^2\theta_c + \sigma_{\mathcal{U}}^2\sin^2\theta_c + \sigma_{\mathcal{U}\mathcal{Q}}^2\sin 2\theta_c}\, .
\end{equation}

\section{Complex Clean applied to Linear Polarisation Images}
\begin{figure*}
\centering
    \includegraphics[width=\textwidth, trim=0.8cm 0 0 0]{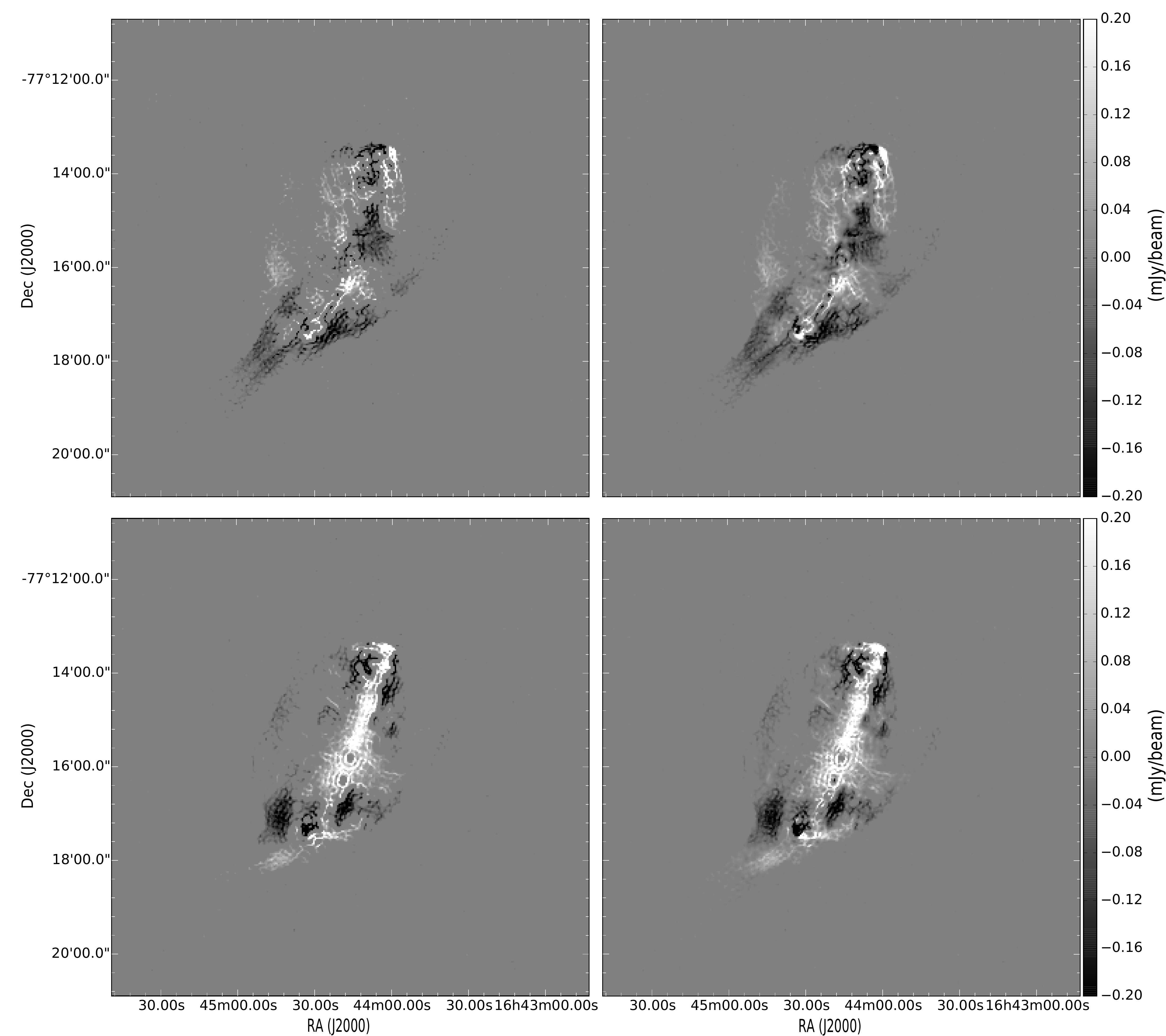}
	\caption{Stokes $\mathcal{Q}$ (top row) and Stokes $\mathcal{U}$ (bottom row) CLEAN component images using the standard (left column) and Complex (right column) SDI CLEAN methods. The Complex SDI CLEAN creates more components related to the smooth structure of the source, while the standard SDI CLEAN creates more CLEAN components associated with the noise.}
    \label{fig:1637pc}
\end{figure*}

\begin{figure*}
\centering
    \includegraphics[width=\textwidth, trim=0.8cm 0 0 0]{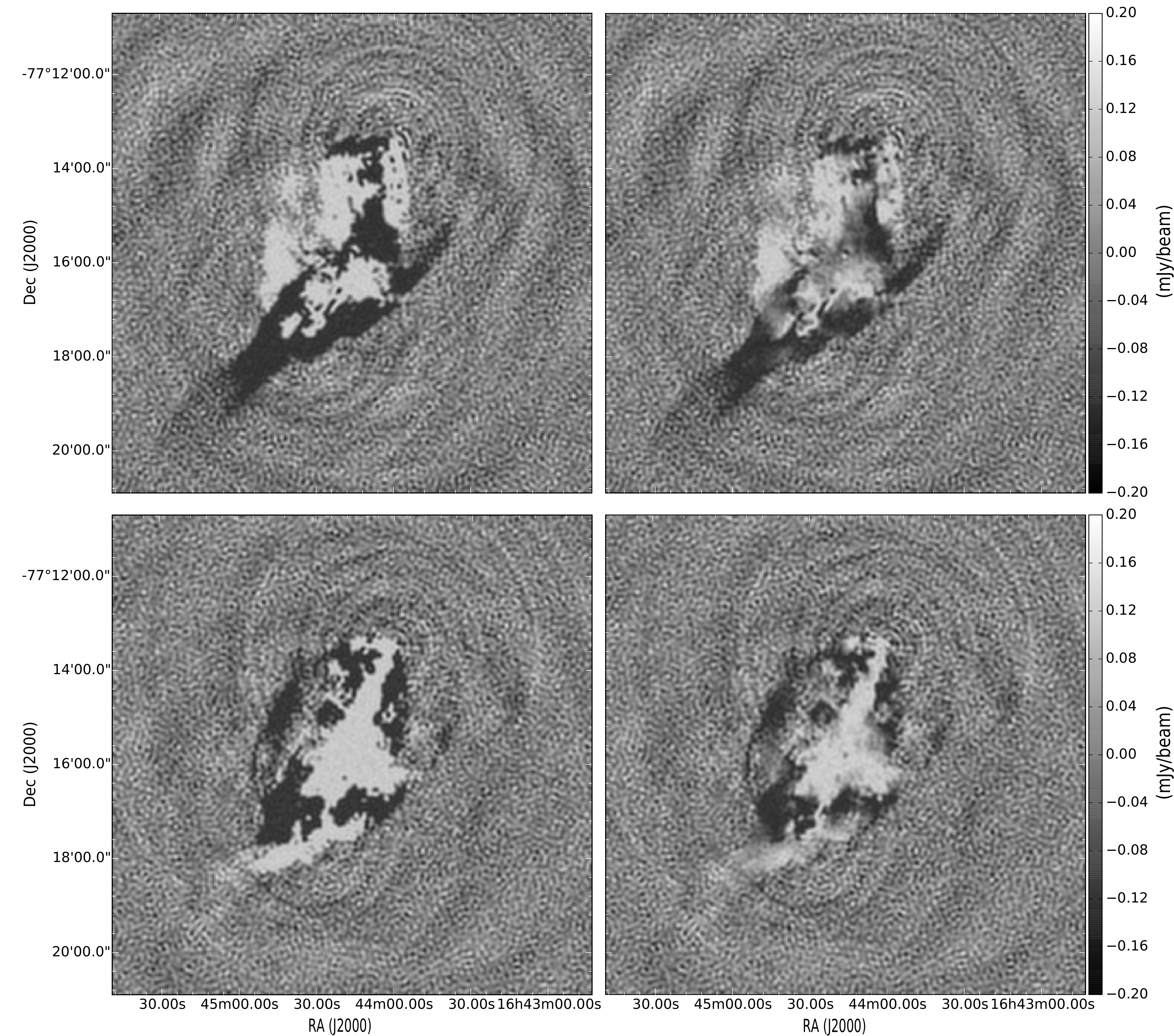}
	\caption{Stokes $\mathcal{Q}$ (top row) and Stokes $\mathcal{U}$ (bottom row) residual images made using the standard (left column) and Complex (right column) SDI CLEAN methods. The residuals made using the Complex SDI CLEAN are lower than the residuals made using the standard SDI CLEAN. Furthermore, the residuals made using the standard SDI CLEAN are smooth over the positive and negative domains, but discontinuous at the domain boundaries (seen as canals in the polarisation residual intensity). The residuals for the Complex SDI CLEAN are smooth across these boundaries and correctly preserve the overall structure of the source.}
    \label{fig:1637res}
\end{figure*}

\begin{figure*}
\centering
    \includegraphics[width=\textwidth, trim=0.8cm 0 0 0]{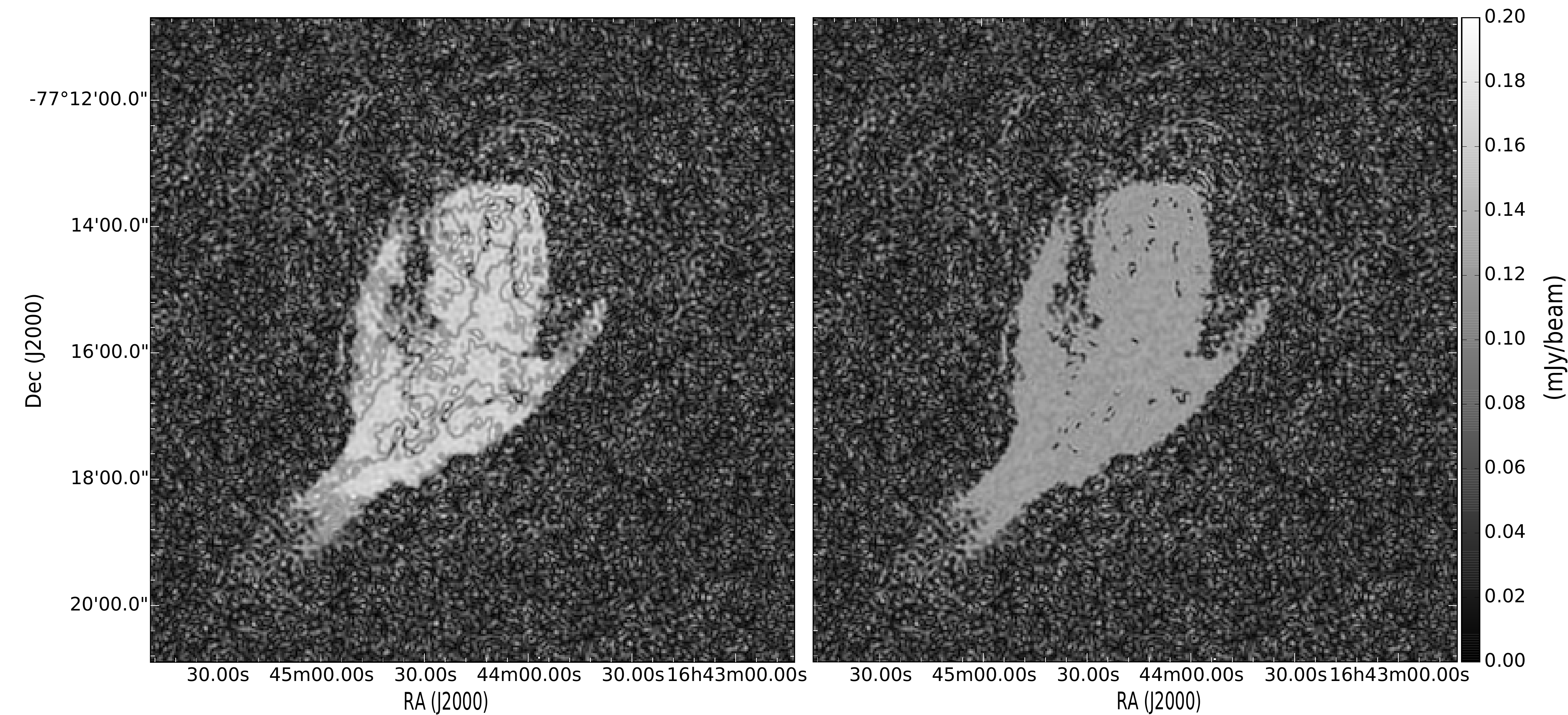}
	\caption{Linear polarisation residual intensity image of PKS B1637-771 using made the standard (left) and Complex (right) SDI CLEAN methods. The complex method shows a smooth residual, and the standard method shows canals within the residuals. These canals are nulls in the residuals of $\mathcal{Q}$ and $\mathcal{U}$, due to a change in sign.}
    \label{fig:1637pres}
\end{figure*}

\begin{figure*}
\centering
    \includegraphics[width=\textwidth, trim=0 0 0 0]{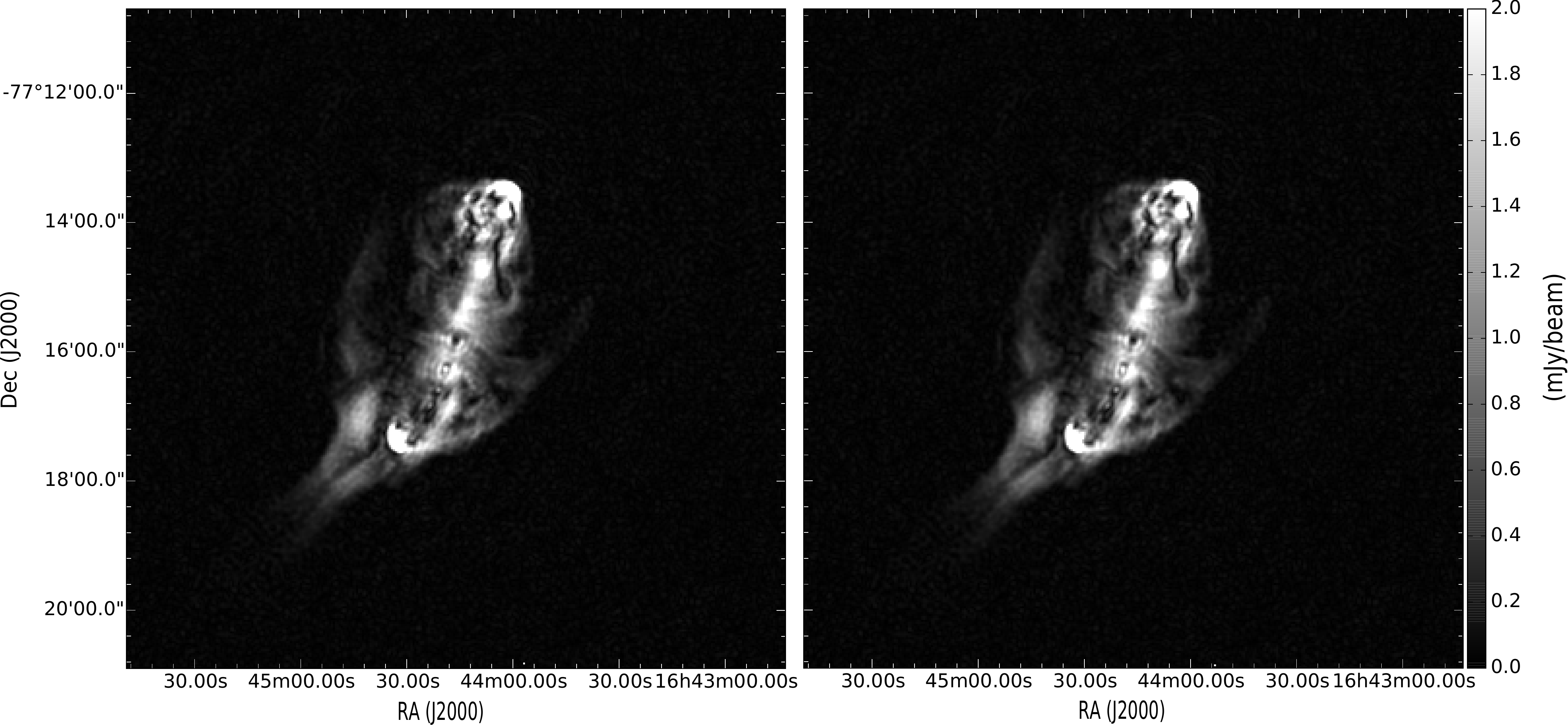}
	\caption{Total linear polarisation intensity images of PKS B1637-771 made using the standard (left) and Complex (right) SDI CLEAN methods. }
    \label{fig:1637pi}
\end{figure*}

Having demonstrated the rotational invariance of Generalised Complex CLEAN we now present a case study to highlight the advantages of the process over the traditional methods by examining the images which result from a standard versus Complex SDI CLEAN of the PKS B1637-771 data. The Stokes $\mathcal{Q}$ and $\mathcal{U}$ CLEAN component images for the standard and Complex SDI CLEAN are shown in Figure \ref{fig:1637pc} while the respective residual images are shown in Figure \ref{fig:1637res}. As shown in the figures the Complex SDI clean finds more components associated with the diffuse structure in the source and almost no components outside the source. By comparison the standard method misses many components associated with the diffuse emission and inserts spurious components associated with noise. This is consistent with the differences in accessibility of components in the $\mathcal{Q}\mathcal{U}$ plane discussed in Section \ref{sec:cutoff}. Additionally, for the Generalised Complex CLEAN we see the residual values change smoothly from positive to negative through a range of values correctly representing the source structure, as opposed to effectively being single valued with a hard transition between the positive and negative domains.  These hard transitions between domains result in the so-called canals (similar to `depolarisation' canals') seen in the residual image of the polarised intensity for the standard CLEAN shown in the left panel of Figure \ref{fig:1637pres}. Furthermore, as these features are a result of the rotational invariance the canals and domains change with rotations of linear polarisation before deconvolution. By comparison the residual image for the total linear polarisation derived from the Generalised Complex CLEAN shown in the right panel of Figure \ref{fig:1637pres} is both smooth and shows less remaining flux. The final restored total linear polarisation images for both methods are shown in Figure \ref{fig:1637pi}, though it is difficult to tell in the greyscale image more diffuse polarised flux is recovered by the Generalised Complex CLEAN and less iterations are required. 

\section{Distribution of Code}

Python code for Complex H\"{o}gbom and SDI CLEANs are available from the authors. Additionally, we have written a patch to allow a Complex SDI CLEAN task to run within MIRIAD. The patch is also available from the authors and we hope that CSIRO will consider including it in the standard MIRIAD release in the future. At present the method is designed to work for single pointing data, however, we have formulated an extension of the Complex SDI CLEAN for mosaiced data (see Appendix) which may be incorporated in MIRIAD in the future.

\section{Conclusions}
In this work, we have presented a new method of CLEANing polarimetric data. We show that our method, `Generalised Complex CLEAN', provides a more physically consistent CLEAN model than the currently used methodologies, particularly for complex, diffuse sources. Generalised Complex CLEAN overcomes many of the problems with current CLEAN by properly accounting for the complex vector nature of polarisation. By looking for peaks in $\mathcal{P}$ instead of separately in $\mathcal{Q}$ and $\mathcal{U}$, Generalised Complex CLEAN provides a method that is rotationally invariant and thus detects more true components at low sign-to-noise and fewer spurious ones and does not suffer from canals in the residual images, thereby recovering more flux. Furthermore, we show that by using Generalised Complex CLEAN the magnitude of the CLEAN components is more accurate than for standard CLEAN algorithms. By examining the relationship between the CLEAN components generated in a rotated and non-rotated frame we are able to measure the effects of numerical precision for different CLEAN methods. We find that the Complex SDI method is considerably better than a Complex H\"{o}gbom CLEAN for preserving properties of components under rotations. This is believed to be the result of less numerical error in the SDI process. If this is the case, then other deconvolution methods which use fewer iterative subtractions may also be more suitable for large images. Both the Python code and a MIRIAD patch to run Generalised Complex CLEAN algorithms will be made available to the community.

Data from a number of current centimetre-wavelength instruments should benefit from the application of Generalised Complex CLEAN including the ATCA, SMA, ALMA, JVLA, GMRT and WRST. Furthermore, as we move to higher resolutions with instruments such as the the Square Kilometre Array which will generate more and more CLEAN components, it will be vital to consider the effects of cumulative numerical errors. In particular SKA1-MID, which will operate in the centimetre wavelength regime, is expected to generate images with of order 35,000 $\times$ 35,000 pixels requiring vast numbers of CLEAN components. Considering the results presented here it would seem prudent for future instruments to adopt a Complex SDI CLEAN in order to both have more physically meaningful results and reduce cumulative numerical errors. At meter-wavelengths instruments like SKA1-Low, which will produce wider and deeper images than ever before with of order 80,000 $\times$ 80,000 pixels \citep{dehghan16}, will certainly suffer from cumulative errors if deconvolution processes with large numbers of iterative steps are used. Such low frequency instruments may also benefit from deconvolution via a Complex SDI clean, but the issues of direction dependent calibration, and primary and synthesized beam variations across the large fields of view make this a more complex and challenging and case. Thus, more work will be needed in these areas to assess the advantages of a Generalised Complex CLEAN on widefield, low frequency data.

% unnumbered section
\section*{Acknowledgements}
MJ-H and LP acknowledge support from the Marsden Fund administered by the Royal Society of New Zealand. We thank Dr Michiel Brentjens and Dr Andre Offringa for clarifying the CLEANING approach and possibilities in CASA and WSClean. We also acknowledge Dr Roman Klapaukh for input to preliminary discussions on the computation impact of the complex deconvolution. Finally we thank the referee and the editors for their constructive suggestions which helped expand the applicability of this work.

\bibliographystyle{mn2e}
\bibliography{paper_r2_arXiv}

%%%%%%%%%%%%%%%%% APPENDICES %%%%%%%%%%%%%%%%%%%%%

\appendix
\section{Complex SDI CLEAN adapted for mosaics}

\citet{sault96} explained that only minimal changes are needed to extend SDI CLEAN to operate on mosaics. Similarly, Complex SDI CLEAN can be extended to deconvolve polarisation mosaics. In this section, we describe the minimal changes needed to extend complex SDI CLEAN to mosaics.

In particular, for mosaicing the residual is weighted by the RMS image of the image when locating the peak. We adopt the same approach, and define the linear polarisation RMS image as $\sigma_\mathcal{P}(x,y)$, where $\sigma_\mathcal{P}(x,y)$ is the RMS of the complex linear polarisation at $(x,y)$, calculated using the variance
\begin{equation}
\sigma_\mathcal{P}^2 = \frac{1}{\ell}\sum_{k=1}^\ell \left | \mathcal{P}(x_k,y_k) - \mu_\mathcal{P}\right |^2 \, ,
\end{equation}
and $\mu_\mathcal{P}$ is the mean in complex linear polarisation and $\ell$ is the total number of samples used to calculate the variance.

First the maximum of $\frac{\left| \mathcal{R}_\mathcal{P}\right|}{\sigma_\mathcal{P}}$ is found, then $\mathcal{M}$ is constructed by clipping $\mathcal{R}_\mathcal{P}$ for values $\frac{\left| \mathcal{R}_\mathcal{P}\right|}{\sigma_\mathcal{P}}<\frac{\alpha\left| \mathcal{R}_\mathcal{P}(x_m,y_m)\right|}{\sigma_\mathcal{P}(x_m,y_m)}$.

The final difference is the calculation of $\eta$, the factor taking into account the scaling needed for the synthesized model $\mathcal{M}_\mathcal{D}$. This is now calculated with weighting each image by the RMS
\begin{equation}
\tilde{\eta} = \frac{\sum_{k=1}^N \mathcal{R}_\mathcal{P}(x_k,y_k)\mathcal{M}_D^\star(x_k,y_k)/\sigma_\mathcal{P}(x_k,y_k)^2}{\sum_{k=1}^N \mathcal{M}_D(x_k,y_k)\mathcal{M}_D^\star(x_k,y_k)/\sigma_\mathcal{P}(x_k,y_k)^2}\,.
\end{equation}
Then, one can scale the synthesized model by $\eta = {\rm max}\left\{0.02,| \tilde{\eta} |\right\}\frac{\tilde{\eta}}{| \tilde{\eta} |}$.

All other steps and calculations of the complex SDI method remain the same.
\footnote{The RMS image $\sigma_\mathcal{P}(x,y)$ cannot be generated from $\sigma_\mathcal{Q}(x,y)$ and $\sigma_\mathcal{U}(x,y)$ because $\sigma_\mathcal{P}(x,y)^2 \neq \sigma_\mathcal{Q}(x,y)^2 + \sigma_\mathcal{U}(x,y)^2$. However, MIRIAD does not have the function for calculating $\sigma_\mathcal{P}(x,y)$, so $\sigma_\mathcal{Q}(x,y)^2 + \sigma_\mathcal{U}(x,y)^2$ may be used as an approximation.}

%%%%%%%%%%%%%%%%%%%%%%%%%%%%%%%%%%%%%%%%%%%%%%%%%%

\end{document}